\documentclass{svjour3}
\usepackage{graphicx}
\usepackage{float}
\usepackage[justification=centering]{subfig}
\usepackage{cite}
\usepackage{hyperref}
\hypersetup{hidelinks}
\usepackage{url}
\usepackage{listings}
\usepackage{xcolor}
\usepackage{tcolorbox}
\usepackage{caption}
\usepackage{booktabs}
\usepackage{multirow}
\usepackage{enumerate}
\usepackage{textcomp}
\usepackage{booktabs}
\usepackage[sort,compress]{natbib}
\usepackage{diagbox}
\usepackage{enumitem}
\usepackage[misc,geometry]{ifsym}
\usepackage{titlesec}
\usepackage{color}
\usepackage{fontawesome}

\usepackage{longtable,tabu}
\usepackage{afterpage}
\usepackage{verbatim}
\usepackage{xurl}
\usepackage{tikz} 
\usepackage{rotating}
\usepackage[toc,page]{appendix}
\usepackage{graphicx}
\usepackage[T1]{fontenc}
\usepackage{pdflscape}
\usepackage{subfloat}
\usepackage{longtable}
\usepackage{ltxtable}
\usepackage{fancyvrb}
\usepackage{tabularx}

\newtcolorbox{qoutebox}[3][]
{
  colframe = gray!30!white,
  colback  = #2!10,
  #1,
}

\begin{document}

\title{Mining Architectural Information: A Systematic Mapping Study}
\titlerunning{Mining Architectural Information}

\author{Musengamana Jean de Dieu \and
        Peng Liang      \and
        Mojtaba Shahin  \and
        Chen Yang       \and
        Zengyang Li     
}

\institute{Musengamana Jean de Dieu \and Peng Liang (\Letter) \at School of Computer Science, Wuhan University, Wuhan, China \\ 
            Hubei Luojia Laboratory, Wuhan, China\\
            \email{mjados@outlook.com, liangp@whu.edu.cn}
            \and
            Mojtaba Shahin \at School of Computing Technologies, RMIT University, Melbourne, Australia \\ 
            \email{mojtaba.shahin@rmit.edu.au}
            \and
            Chen Yang \at School of Artificial Intelligence, Shenzhen Polytechnic University, Shenzhen, China \\ 
            State Key Laboratory for Novel Software Technology, Nanjing University, Nanjing, China \\
            \email{yangchen@szpu.edu.cn}
            \and 
            Zengyang Li \at School of Computer Science \& Hubei Provincial Key Laboratory of Artificial Intelligence and Smart Learning, Central
China Normal University, Wuhan, China \\ 
            \email{zengyangli@ccnu.edu.cn}
}

\date{Received: date / Accepted: date}

\maketitle

\begin{abstract}
Mining Software Repositories (MSR) has become an essential activity in software development. Mining architectural information (e.g., architectural models) to support architecting activities, such as architecture understanding, has received significant attention in recent years. However, there is a lack of clarity on what literature on mining architectural information is available. Consequently, this may create difficulty for practitioners to understand and adopt the state-of-the-art research results, such as what approaches should be adopted to mine what architectural information in order to support architecting activities. It also hinders researchers from being aware of the challenges and remedies for the identified research gaps. We aim to identify, analyze, and synthesize the literature on mining architectural information in software repositories in terms of architectural information and sources mined, architecting activities supported, approaches and tools used, and challenges faced. A Systematic Mapping Study (SMS) has been conducted on the literature published between January 2006 and December 2022. Of the 104 primary studies finally selected, 7 categories of architectural information have been mined, among which architectural description is the most mined architectural information; 11 categories of sources have been leveraged for mining architectural information, among which version control system (e.g., GitHub) is the most popular source; 11 architecting activities can be supported by the mined architectural information, among which architecture understanding is the most supported activity; 95 approaches and 56 tools were proposed and employed in mining architectural information; and 4 types of challenges in mining architectural information were identified. This SMS provides researchers with promising future directions and help practitioners be aware of what approaches and tools can be used to mine what architectural information from what sources to support various architecting activities.

\end{abstract}
\section{Introduction} \label{introduction} 
Software repositories contain a wealth of valuable information (e.g., source code, bug reports) related to software development. This information can assist software engineers (e.g., architects and developers) to comprehend various aspects of a software system. Architectural information is one of the most important types of information in software development, and it is not only used in the early stages (e.g., architecture design) of the development, but also in the later stages of software development life cycle, like maintenance and evolution \citep{li2013application}. Architectural information, such as benefits and drawbacks of certain architectural solutions (e.g., patterns and tactics) in specific application domains, can help architects and related stakeholders to conduct architecture evaluation on candidate architectural solutions during the architecting process \citep{hofmeister2007general}. However, architectural information is scattered in various sources of software repositories (e.g., Q\&A sites \citep{soliman2016architectural}, technical blogs and tutorials \citep{soliman2021exploring}, issue tracking systems \citep{soliman2021exploratory}, developer mailing lists \citep{ding2015understanding}, and chat messages \citep{borrego2019towards}), and is often described in a combination of textual and graphical representation \citep{Malavolta2013WhatIN}. This constitutes a large volume of architectural information that is sometimes tacit and not documented at all in those sources \citep{ding2014open}. Therefore, the intricate information of a system, especially in a large and complex system easily evaporates if architectural information is non-documented. The consequences would be incurring design and implementation issues \citep{Capilla201610YO}. With the increase in the complexity and size of systems, architectural information management becomes even more challenging, and software engineers need to find the right information and recover it efficiently. However, searching and finding or recovering relevant architectural information from artifacts (e.g., code, requirements document) is a challenging task for software engineers \citep{JANSEN20091232}. Therefore, approaches and tools for searching and mining architectural information from development artifacts are much needed to assist the development process (e.g., architecting process). 

The Mining Software Repositories (MSR) field analyzes the rich data available in software repositories to uncover interesting and actionable information about software systems \citep{hassan2008road}. Mining software repositories for Software Architecture (SA) has been the subject of the software architecture research community in recent years \citep{soliman2021preface}. SA researchers and practitioners have developed approaches and tools to ease the mining of architectural information from various sources of software repositories in order to support architecting activities \citep{li2013application, weinreich2012towards}. Some of the supported architecting activities include architecture analysis \citep{velasco2016knowledge}, architecture understanding \citep{do2015keecle}, and architecture recovery \citep{Shahbazian2018RecoveringAD}. Due to the increasing importance of mining architectural information, a growing number of researchers have explored different aspects of this topic. Despite this, there is a lack of deep understanding of current literature on mining architectural information, which hinders researchers from being aware of the potential challenges and remedies for the identified research gaps. It also creates difficulty for practitioners to understand and adopt the state-of-the-art research results, such as what approaches should be adopted to mine what architectural information in order to support architecting process. Thus, this motivated us to carry out a literature review of the reported research on mining architectural information from software repositories to support architecting activities.


(1) We presented a comprehensive analysis of 104 primary studies published between 2006 and 2022 on mining architectural information in software development.

(2) We provided a classification of mined architectural information to support the development.

(3) We presented the sources used in mining architectural information.

(4) We investigated the architecting activities that can be supported by the mined architectural information.

(5) We identified the approaches and tools that have been used in mining architectural information. 


The remainder of this paper is organized as follows: Section \ref{researchContext} introduces the context of this SMS. Section \ref{MappingDesign} elaborates on the design and research questions of this SMS. Section \ref{Results} presents the results of each research question. Section \ref{Discussion} discusses the results of the research questions and their implications for researchers and practitioners. Section \ref{ThreatValidity} examines the threats to validity. Section \ref{RelatedWork} presents the related work. Finally, Section \ref{ConclusionFurtureWork} concludes this SMS with potential areas of future research.

\section{Research Context} \label{researchContext}
To clarify the scope of this SMS, two fundamental concepts need to be explained, mining software repositories and architectural information.

\subsection{Mining Software Repositories}
The studies on mining software repositories collect and analyze the rich data available in software repositories to uncover interesting and actionable information about software projects \citep{hassan2008road}. VCS (e.g., GitHub), Q\&A sites (e.g., Stack Overflow), and issue tracking systems (e.g., Jira), among others, are examples of sources of software repositories that are commonly used for mining development related information. These sources are useful for both software practitioners and researchers. For example, practitioners (e.g., architects) can share and learn architectural tactics and quality attributes knowledge from their peers in Stack Overflow \citep{bi2021mining}. Researchers can benefit from the available data (e.g., source code, commit messages) in software repositories to conduct their research on software development, such as code clone detection \citep{nafi2019clcdsa} and dependency analysis \citep{bavota2014improving}. To achieve this, researchers are required to select software repositories and data sources that fit their research needs, extract data from these repositories, and analyze the data to obtain evidence for answering their research questions. In this SMS, we gather the literature on mining architectural information from software repositories to support the development. Specifically, we search the studies that develop approaches and tools for mining architectural information, aiming to support architecting activities (e.g., mining code to support architecture maintenance and evolution \citep{kazman2015case}).

\subsection{Architectural Information}
Architectural information states a high-level abstraction of software systems, and it is an important type of information in software development \citep{SA2012}. There are various types of architectural information, such as Architecturally Significant Requirements (ASRs), architectural solutions, and architectural decisions that are intensively explored during the development. In this section, we introduce the concept of architectural information through these three concrete types of architectural information.

To create an architecture, one must understand the problem, find an architectural solution, and evaluate the final result \citep{souza2019deriving}. Understanding the problem requires identifying, analyzing, and prioritizing the stakeholders’ needs, producing the so-called \textit{architecturally significant requirements}, such as functional and non-functional requirements \citep{souza2019deriving}. \textit{Architectural solutions} such as architectural patterns (e.g., Model–View–Controller), tactics (e.g., resource pooling), and frameworks (e.g., Django) are the fundamental building blocks in modern software design and they are used to address the ASRs \citep{SA2012}. Finding \textit{architectural solutions} is related to deriving a software architecture from the requirements \citep{souza2019deriving}. However, finding a suitable \textit{architectural solution} from the existing alternative solutions requires a series of design reasoning to satisfy the stakeholders’ needs and the system’s requirements (e.g., performance, security, scalability). Finally, evaluating architectural solutions requires deciding whether, and to what extent, the chosen solutions solve the problems. \textit{Architectural decision} is a description of a set of architectural additions, subtractions, and modifications to the architecture, the rationale, design rules, design constraints, and additional requirements that (partially) realize one or more requirements on a given architecture \citep{jansen2005SoftArch}. \textit{Architectural decisions} play a crucial role in software architecture, during the design, implementation, evolution, reuse, and integration of architectures \citep{jansen2005SoftArch}. 

Architectural information is often described in different formats, such as textual and graphical representation \citep{Malavolta2013WhatIN}, and this information is recorded in various artifacts, such as code, architecture documents \citep{ding2014open}, that are contained in diverse sources of software repositories. However, it is very challenging for architects and developers to mine and reuse architectural information that is contained in those artifacts due to, for instance, unstructured representation of this information in texts or graphs. As mentioned in Section \ref{introduction}, mining repositories for SA has been the subject of the architecture research community in recent years \citep{soliman2021preface}. Researchers and practitioners have developed many approaches and tools to facilitate the mining of architectural information from various artifacts that are contained in software repositories in order to support the development. 
\section{Mapping Study Design} \label{MappingDesign}
In this study, we decided to conduct an SMS over other types of secondary studies, such as Systematic Literature Review (SLR) or survey. The reason is that SMS provides an overview of a research area by systematically identifying and evaluating the evidence in literature. One of the main differences between an SMS and an SLR is that an SMS aims to discover the research trends and covers a broad topic in the literature, while an SLR usually has a relatively narrow and in-depth research scope and focuses on specific research questions \citep{kitchenham2007guidelines}.

\subsection{Goal and Research Questions}\label{GoalResearchQuestions}
The goal of this SMS is to \textit{\textbf{analyze}} the primary studies on mining architectural information \textit{\textbf{for the purpose of}} understanding \textit{\textbf{with respect to}} the sources of architectural information and types of mined architectural information. This SMS also investigates the supported architecting activities, approaches, and tools employed as well as challenges \textit{\textbf{from the point of view of}} researchers \textit{\textbf{in the context of}} software development. In order to get a comprehensive overview of mining architectural information in software development, we further decomposed the goal of this SMS into five Research Questions (RQs) as listed in Table~\ref{ResearchQuestions}.

\begin{table}
\small
	\caption {Research questions and their rationale}
	\label{ResearchQuestions}
        \resizebox{\columnwidth}{!}{
	\begin{tabular}{p{5cm}p{10.5cm}}
		\toprule
	\textbf{Research Question} & \textbf{Rationale}\\
		\midrule
	   \textbf{RQ1.} What architectural information is mined in software development? & SA researchers have mined different types of architectural information, such as architectural description \citep{chaabane2017mining} and system requirements (e.g., quality attributes \citep{bi2021mining}). This RQ intends to explore the types of mined architectural information to support software development. The answer of this RQ can (i) help practitioners know and utilize the types of architectural information that have been mined to support the development and (ii) help researchers be aware of the types of architectural information that have and have not been mined so that they can explore how to use the mined architectural information and/or mine the architectural information that is not yet mined.\\
	 
	  \textbf{RQ2.} What sources are used to mine architectural information? & Architectural information is scattered in various sources of software repositories, such as Q\&A sites \citep{soliman2016architectural}, issue tracking systems \citep{bhat2017automatic}, and technical blogs and tutorials \citep{de2022developers}. There is no single source of architectural information that contains all architectural information concepts, and certain architectural information concepts come more often from specific architectural information sources \citep{soliman2021exploring}. 
      For example, developer communities (as an example of architectural information source) contain predominantly general architectural information concepts, such as the benefits and drawbacks of architectural solutions \citep{soliman2017developing}. In contrast, issue tracking systems contain mainly architectural issues and design decisions of existing systems \citep{bhat2017automatic}. On the other hand, architectural solutions and rationales (e.g., benefits and drawbacks of architectural solutions) are less present within source code repositories, such as GitHub \citep{soliman2021exploring}. 
      Therefore, researchers have mined various sources for the sake of certain architectural information to support the development. The answer of this RQ provides insights into specific sources researchers utilize when mining architectural information, and such insights can (i) motivate other researchers to further investigate architectural information in those sources and (ii) help practitioners know where they can search architectural information to address their architectural design concerns.\\
	 
	  \textbf{RQ3.} What architecting activities can be supported by the mined architectural information? & There are various architecting activities (e.g., architecture understanding, architecture description, architectural implementation, architecture maintenance, architecture recovery \citep{li2013application, weinreich2012towards}) that are performed during the architecture life cycle. Each architecting activity may require specific architectural information to be conducted effectively. The answer of this RQ can help practitioners know which architecting activities can be supported by the mined architectural information. For example, during architecture evaluation, architectural information, such as the benefits and drawbacks of certain architectural solutions (e.g., architectural patterns and tactics) can help architects evaluate the architectural solutions (e.g., choosing suitable architectural patterns according to the positively and negatively affected quality attributes) in specific application domains.\\
	 
     \textbf{RQ4.} What approaches and tools are used to mine architectural information? & There might be approaches and tools that have been proposed and used to mine architectural information. The answer of this RQ can provide practitioners with an overview of the readily available approaches and tools that they can utilize to mine architectural information from software repositories, and to some extent inspire researchers to develop new and dedicated architectural information mining approaches and tools.\\
	 
     \textbf{RQ5.} What are the challenges in mining architectural information? & Mining architectural information in software repositories has many challenges, such as vagueness or ambiguity in architectural element description and applicability of approaches to heterogeneous architectural datasets. With this RQ, we want to identify and report these challenges as potential future research directions in this area.\\
		\bottomrule 
	\end{tabular}}
\end{table}

\subsection{Mapping Study Execution}
\begin{figure}
 \centering
  \includegraphics[width=1.0\linewidth]{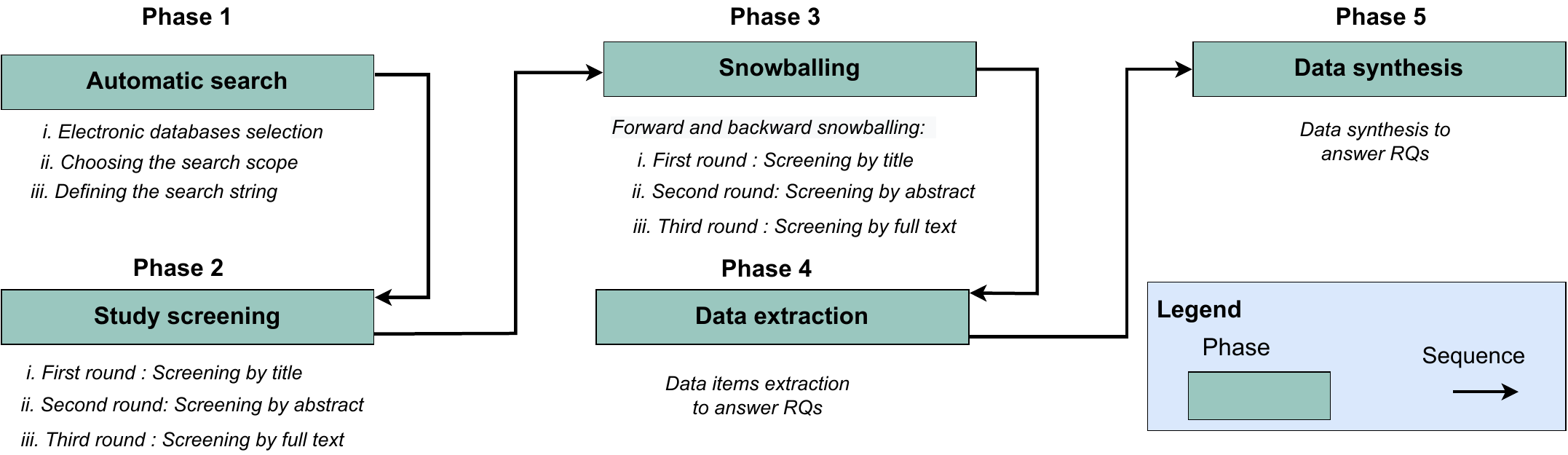}
 \caption{Mapping study execution}
 \label{MappingStudyExecution}
\end{figure}

We designed this SMS according to the guidelines for systematic mapping studies proposed by \cite{petersen2015guidelines}. Figure \ref{MappingStudyExecution} illustrates the five phases (i.e., automatic search, study screening, snowballing, data extraction, and data synthesis) that we followed to execute this mapping study.

\subsubsection{Phase 1: Automatic Search}\label{AutomaticSearch}
We performed the automatic search in eight electronic databases (see Table \ref{ElectronicDatabases}) with three steps (i.e., electronic databases selection, choosing the search scope, and defining the search string). 

     \textit{(i) Electronic databases selection}: The selected electronic databases are shown in Table \ref{ElectronicDatabases} and 
     they are regarded as the most common and popular databases in the software engineering field to search for relevant studies \citep{chen2010towards}. Google Scholar was not included in this SMS because through a pilot search we observed that Google Scholar produced a significant number of irrelevant studies and the results from Google Scholar were overlapped with the studies returned from other databases.
       
     \textit{(ii) Choosing the search scope}: We set the starting date to January 2006. The starting year is justified by considering the milestone paper about the golden age of software architecture in research and practice that was published in 2006 by \cite{shaw2006golden}. The ending search period is settled as December 2022 when we started this SMS.
     
     \textit{(iii) Defining the search string}: To define the search string used in this SMS, we followed the following steps: 
     (1) We extracted the major terms (e.g., “architect*”, “mining software repositor*”, “Stack Overflow”) based on the research topics and research questions in the existing relevant studies (e.g., \cite{soliman2021exploring, malavolta2021mining}) that were already known to us. (2) We generated a list of synonyms (i.e., topic-related terms) for each extracted major term. For example, the synonym terms for “architect*” term are “design*” and “structur*”. (3) To compose the search string, we combined each extracted major term and its synonyms with the Boolean operator \textbf{OR}, and (4) we linked the major terms with the Boolean operator \textbf{AND}. 
     
     Note that, before the formal search and selection, we conducted a pilot search and selection with an initial set of search terms in order to decide the search terms used in the formal search and selection. The pilot search can help settle the appropriate search terms for the formal search \citep{petersen2015guidelines}. Based on the pilot search and selection results, the search terms were adjusted and refined accordingly. Alternative synonyms and spellings for search terms were then modified. These procedures provided a high confidence that the majority of the primary studies were identified. To clarify, two search strings were used in this pilot search, i.e., (“\textit{architecture}” OR “\textit{design}” OR “\textit{structure}”) AND (“\textit{mining software repository}” OR “\textit{repository mining}”) and (“\textit{architecting}” OR “\textit{designing}” OR “\textit{structuring}”) AND (“\textit{mining software repository}” OR “\textit{repository mining}”). We noticed that the search results by different search terms are complementary to each other, for example, using the term “architecture” in IEEE Explore did not return all relevant studies that were retrieved by using the terms “architecting”, “designing”, and “structuring”. Therefore, we decided to include all these terms in the final construction of the search string to avoid missing potentially relevant studies. Moreover, it should be noted that we did not include the terms related to specific types of architectural information, such as architecture decision, quality attribute requirements, architecture smell, in the search due to two reasons: (1) part of the goal of this SMS is to identify the types of mined architectural information in software development (i.e., RQ1), therefore, we cannot have a relatively comprehensive list of those types before conducting this SMS; (2) including the terms related to the specific types of architectural information in the search query may lead to the situation that the search results are biased to these types. Furthermore, some electronic databases have their own limitations on the number of search terms and operators. For example, IEEE Explore does not allow more than seven wildcards (*) and ScienceDirect does not support more than eight Boolean connectors. Hence, the final search string was adjusted according to the restrictions and settings of each database. Overall, the final search string used in the formal search was defined as: 
    

\noindent\resizebox{\columnwidth}{!}{\fbox
     {%
	\parbox{\columnwidth}{%
		(“\textit{mining software repositor*}” OR “\textit{repositor* mining}”) AND (\textit{architect* OR design* OR structur*}) AND (“\textit{Stack Overflow}” OR \textit{StackOverflow} OR \textit{GitHub} OR “\textit{open source software}” OR “\textit{open source communit*}” OR “\textit{online developer communit*}” OR “\textit{online communit*}” OR “\textit{online developer forum*}” OR “\textit{online forum*}” OR “\textit{question and answer site*}” OR “\textit{question and answer website*}” OR “\textit{Q\&A site*}” OR “\textit{Q\&A website*}” OR “\textit{mailing list*}” OR \textit{gitter} OR \textit{slack} OR \textit{chat} OR “\textit{issue tracker}” OR “\textit{issue tracking}” OR “\textit{issue management}”)
	}
}}
      
\begin{table} 
\small
\caption {Electronic databases used for the automatic search}
\label{ElectronicDatabases}
\resizebox{\columnwidth}{!}{
\begin{tabular}{p{3cm}p{5.9cm}p{4.9cm}}
	\toprule
	    \textbf{Database}       & \textbf{URL}                             & \textbf{Search Scope in Database}\\
		\midrule
	ACM Digital Library     &\url{https://dl.acm.org/}                     & Study title, abstract\\
        IEEE Explore            &\url{https://ieeexplore.ieee.org/}            & Study title, abstract, keywords\\ 
        Science Direct          &\url{https://www.sciencedirect.com/}          & Study title, abstract, keywords \\ 
        EI Compendex            &\url{https://www.engineeringvillage.com/}     & Study title, abstract\\ 
        Springer Link           &\url{https://link.springer.com/}              & Study title, abstract \\
        Wiley InterScience      &\url{https://onlinelibrary.wiley.com/}        & Study title, abstract\\
        ISI Web of Science      &\url{https://login.webofknowledge.com/}       & Study title, abstract, keywords\\ 
        Scopus                  &\url{https://scopus.com/}                     & Study title, abstract, keywords\\ 
		\bottomrule 
	\end{tabular}}
\end{table}

\subsubsection{Phase 2: Study Screening} \label{StudyScreening}
Figure \ref{SearchAndScreeningResults} shows the study screening process. The automatic search in the eight electronic databases resulted in 24,205 potentially relevant studies, where 4,386 studies of them were duplicated. After removing the duplicates, 19,588 studies were retained, and we applied the inclusion and exclusion criteria defined in Table \ref{InclusionExculusionCriteria} to screen the remaining studies (i.e., 19,588 studies). Before the formal study screening (manual inspection), to reach an agreement about the inclusion and exclusion criteria (see Table \ref{InclusionExculusionCriteria}), a pilot study screening was performed whereby the first two authors randomly selected 100 primary studies from 19,588 studies and checked them independently. Specifically, this study screening process involved the following rounds and the inclusion and exclusion criteria elaborated in Table \ref{InclusionExculusionCriteria} were applied in each round: (1) In the first round of study screening, the first two authors independently screened the 100 studies by reading the titles. Any uncertain studies (i.e., they could not be decided by the titles) were temporarily included and kept in the second round. (2) In the second round of study screening, the first two authors independently screened the studies left in the first round screening by reading their abstracts. Those studies were retained for the third round for which the first two authors had not made any decision. (3) In the third round of study, the first two authors independently screened studies by reading the full texts of the studies left in the second round. Moreover, the first two authors held a meeting to compare the pilot study screening results and then discuss their disagreements (if any) in an effort to reconcile them and arrive at a final version in which as many discrepancies as possible have been resolved. To measure the inter-rater agreement between the first two authors, we calculated the Cohen’s Kappa coefficient \citep{cohen1960coefficient} and got an agreement of 0.935. During the formal study screening process, we followed the same rounds of study screening, which were used during the pilot study screening process, and they are described below. The selected number of studies in each round of the study screening process is provided in Figure \ref{SearchAndScreeningResults}. 

     \textit{(i) First round of study screening (i.e., reading titles)}: Through the use of the inclusion and exclusion criteria elaborated in Table \ref{InclusionExculusionCriteria}, the first author screened the studies by reading the titles of the remaining studies from the pilot study screening in order to select potential primary studies. Moreover, he recorded the key terms that led to the inclusion/exclusion of a specific study, and those terms were further used for discussion during the consensus meetings and reassessment process. In order to mitigate potential personal bias during study screening, the results of the first round were checked and validated by the second author. In some cases, where the relevance of a study was still unclear from the title, the first and second authors decided to temporarily include it to the second round of study screening. 
   
     \textit{(ii) Second round of study screening (i.e., reading abstracts)}: In the second round of study screening, the first author read the abstracts of the retained studies from the first round, and he screened the studies based on the inclusion and exclusion criteria (in Table \ref{InclusionExculusionCriteria}). He followed the same procedure (i.e., recording the key terms that led to inclusion/exclusion of a specific study) that was employed in the previous round study screening. Similar to the first round screening, to mitigate potential bias, the second author checked and validated the results from the second round. Moreover, if any disagreements (i.e., whether a study could be included or not) between the two authors arose, such conflicts were discussed and resolved among the first two authors. Those studies were kept for the final round which were hard to decide based on their abstracts. 
     
     \textit{(iii) Third round of study screening (i.e., reading full texts)}: The first author read the full text of each study that was retained from the second round, and he screened the studies based on the selection criteria (in Table \ref{InclusionExculusionCriteria}). He followed the same procedure (i.e., recording the key terms that led to inclusion/exclusion of a specific study, and validating the study screening results with the second author) that was used in the previous rounds. Note that, we got controversies on 11 primary studies from the study screening results (i.e., whether a study should be excluded or included in the final round of study screening). Such controversies between the first two authors were discussed in meetings involving the third author till a consensus was reached. 

\begin{table} [h!]
\small
\captionsetup{font=scriptsize}
	\caption{Inclusion (I) and Exclusion (E) criteria for selecting the primary studies}
	\label{InclusionExculusionCriteria}
       \resizebox{\columnwidth}{!}{
	\begin{tabular}{p{15cm}}
		\toprule
	\textbf{Inclusion criterion}\\ 
	\midrule
     \textbf{I1.} A study that focuses on mining architectural information in software development.\\
	 \midrule
	\textbf {Exclusion criteria}\\
	\midrule
    \textbf{E1.} A study focuses on architectural information without mining it.\\
    \textbf{E2.} A study focuses on mining other types of information instead of architectural information.\\
    \textbf{E3.} A study not written in English is excluded.\\ 
    \textbf{E4.} A short paper (i.e., less than 4 pages) is excluded.\\
    \textbf{E5.} If two studies publish the same work in different venues (e.g., conference and journal), the less mature one is excluded.\\
    \textbf{E5.} A study that is grey literature (e.g., technical report) is excluded. 
		\\ \bottomrule 
	\end{tabular}}
\end{table}

\subsubsection{Phase 3: Snowballing}
\begin{figure} [!h]
 \centering
  \includegraphics[width=1\linewidth]{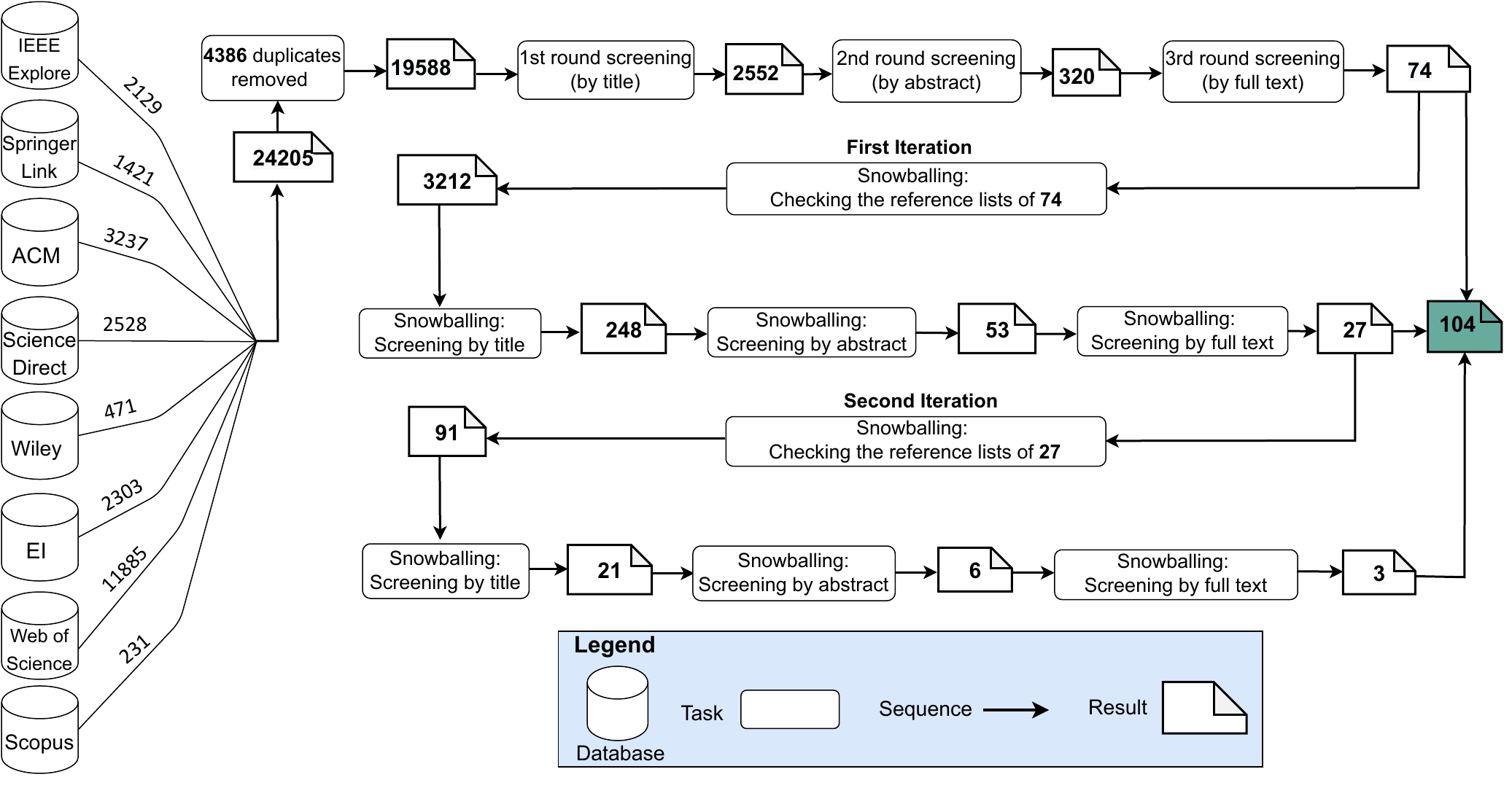}
 \caption{Results of study search and screening}
 \label{SearchAndScreeningResults}
\end{figure}

In order to minimize the threat of missing relevant studies that were possibly missed out during the automatic search, we conducted snowballing by following the approach proposed by \cite{wohlin2014guidelines}. Specifically, we adopted the forward (i.e., collecting those studies citing the selected studies) and backward (i.e., using the references of the selected studies) snowballing. We conducted the snowballing process iteratively, and the iterative process was completed when there were no newly selected studies. Specifically, we performed two iterations during the snowballing process. We used the set of studies selected from the final round of study screening in the previous section (see Section \ref{StudyScreening}) as the start seed (i.e., 74 studies). We then looked for citations from each of the 74 studies in the start seed. We used Google Scholar to collect citations for each paper. We downloaded the citations and analyzed each one according to the inclusion and exclusion criteria defined in Table \ref{InclusionExculusionCriteria}. In the first iteration, 3,228 citations (of the 74 studies) were found. These 3,212 citations were further screened based on the titles (248 studies retained), abstracts (53 studies retained), and full texts (27 studies retained). In the second iteration, we considered a start seed of 27 studies retained from the first iteration and 91 citations were collected. Akin to the citations obtained from the first iteration, the 91 citations were also subjected to the inclusion and exclusion criteria, and 21 studies were retained based on titles, 6 studies were identified based on abstracts, and 3 studies were finally selected based on full texts. Finally, 30 (i.e., 27+3) primary studies were added through the snowballing process. In Figure \ref{SearchAndScreeningResults}, we show the snowballing process that was performed in this SMS. Therefore, we have selected 104 (i.e., 74+30) primary studies in our SMS. The details of the selected 104 primary studies are provided in the Selected Studies file in the Supplementary Material \citep{dataset}. 

\subsubsection{Phase 4: Data Extraction}
To provide the demographic information of the selected studies and answer the five RQs of this SMS (see Table \ref{ResearchQuestions}), we carefully read all the primary studies selected in this SMS (see Figure \ref{MappingStudyExecution}) and extracted the required data items as listed in Table \ref{DataExtraction}. Before the formal data extraction, the first three authors discussed the meaning of each data item and the way about how to extract data. To ensure an unambiguous understanding of the data items, the first author conducted a pilot data extraction. Specifically, the first author randomly selected 15 primary studies (from 104 selected studies, see Figure \ref{SearchAndScreeningResults}) and extracted the data according to the data items in Table \ref{DataExtraction}. 
To suppress the effect of subjective bias, the pilot data extraction results were reviewed and validated by other two authors (i.e., the second and third authors) of this study so that three authors could get a consensus on the understanding of the extracted data items. Likewise, in the formal data extraction process, the data extraction was performed by the first author and validated by the second and third authors, and any divergences and ambiguities of the results in the extracted data were discussed together to reach an agreement. In this way, we can ensure that the extracted data in the formal data extraction process are valid. The description of the data items and their relevant RQs are presented in Table \ref{DataExtraction}. Specifically, nine data items have been defined to be extracted from the selected studies. The three data items (i.e., D1-D3) are used to extract the demographic details of the selected studies, and the remaining data items (i.e., D4-D9) are used to answer the five RQs (see Table \ref{ResearchQuestions}). The data extraction was subsequently followed by data synthesis, and these two processes were conducted and recorded with the aid of MAXQDA (a qualitative data analysis tool).

\begin{table}
\small
	\caption {Data items extracted from the selected studies with the relevant RQs}
	\label{DataExtraction}
        \resizebox{\columnwidth}{!}{
	\begin{tabular}{p{0.3cm}p{2cm}p{5cm}p{5cm}p{1.3cm}}
		\toprule
    \#	&\textbf{Data item}           & \textbf{Description}                                                  & \textbf{Data analysis approach}                   &  \textbf{Relevant RQ}\\
		\midrule
    D1   & Publication year                & The publication year of the study.                               & Descriptive statistics                            & Overview\\       
    D2   & Publication venue               & The name of the venue where the study is published.              & Descriptive statistics                            & Overview\\       
    D3   & Publication type                & The type of the study (i.e., journal, conference, or workshop).  & Descriptive statistics                            & Overview\\ 
    D4   & Architectural information       & The type of mined architectural information.                     & Open coding \& constant comparison, predefined classifications, and descriptive statistics      & RQ1 \\ 
    D5   & Source                          & The source used to mine architectural information.               & Open coding \& constant comparison and descriptive statistics     & RQ2 \\ 
    D6   & Dataset                         & The dataset used to mine architectural information.              & Open coding \& constant comparison and descriptive statistics     & RQ2 \\
    D7   & Architecting activity           & The architecting activity that a study claims to support.        & Open coding \& constant comparison, predefined classifications, and descriptive statistics      & RQ3 \\
    D8   & Approach                        & The approach used to mine architectural information.             & Open coding \& constant comparison and descriptive statistics     & RQ4 \\
    D9   & Tool                            & The tool used to mine architectural information.                 & Open coding \& constant comparison and descriptive statistics     & RQ4 \\ 
    D10  & Challenge                       & The challenge faced when mining architectural information.       & Open coding \& constant comparison and descriptive statistics     & RQ5 \\ 
		\bottomrule 
	\end{tabular}}
\end{table}

\subsubsection{Phase 5: Data Synthesis}\label{DataSynthesis}
During this phase, we synthesized the extracted data that we gathered from the previous phase (i.e., Phase 4: Data extraction) in order to answer the five RQs (see Table \ref{ResearchQuestions}). We used open coding \& constant comparison to analyze several data items and answer certain RQs of this SMS (see Table \ref{DataExtraction}). 
Specifically, in this SMS, open coding \& constant comparison were employed for analyzing the data items D5 and D6 to answer RQ2 (sources), the data items D9 and D9 to answer RQ4 (approaches and tools), and the data item D10 to answer RQ5 (challenges). In addition, descriptive statistics \citep{wohlin2003empirical} was also used for analyzing the data items D5 and D6 to answer RQ2 (sources), the data items D8 and D9 to answer RQ4 (approaches and tools), and the data item D10 to answer RQ5 (challenges). Descriptive statistics can provide quantitative summaries based on the initial description of the extracted data. Moreover, we used a hybrid approach which is a combination of open coding \& constant comparison, predefined classifications (i.e., the conceptual model for architectural description in the ISO 42010:2011 standard \citep{6129467}, the categories of architecture decisions by \cite{kruchten2004ontology}, and the categories of architectural changes by \cite{williams2010characterizing}), and descriptive statistics for analyzing the data item D4 to answer RQ1 (mined architectural information). On another hand, a hybrid approach which is a combination of open coding \& constant comparison, predefined classifications of architecting activities by \cite{hofmeister2007general}, \cite{tang2010comparative}, and \cite{li2013application}, and descriptive statistics was employed for analyzing the data item D7 to answer RQ3 (architecting activities). 

As mentioned above, we utilized a qualitative data analysis tool MAXQDA\footnote{\url{https://www.maxqda.com/}} to support the analysis process. 
Specifically, before the formal data analysis (manual labeling), to reach an agreement about the data items that we gathered from the previous phase (i.e., Phase 4: Data extraction), we first performed a pilot data analysis. Specifically, this analysis process involved the following steps: (1) The first and second authors independently checked and read a random sample of 5 primary studies (from 104 selected studies, see Figure \ref{SearchAndScreeningResults}). (2) The first two authors independently labeled the extracted data with codes that succinctly summarize the data items (see Table \ref{DataExtraction}) for answering RQs. (3) The first two authors independently grouped all the codes into higher-level concepts and turned them into categories or subcategories. The grouping process was iterative, in which each author continuously went back and forth between the code, concepts, categories, and extracted data items to revise and refine both the code, concepts, and categories. In order to improve the reliability of the pilot data analysis results, the first two authors held a meeting and followed a negotiated agreement approach~\citep{campbell2013coding} to compare the coding results, then discussed their disagreements and uncertain judgments on the data encoding results in an effort to reconcile them and arrive at a final version of the pilot data analysis results in which all the discrepancies have been resolved. 
The first author carried on with the formal data analysis and followed the same steps used during the pilot data analysis. In the following paragraph, we provide the details of the formal data analysis process.

The first author fully read the full text of each selected study from the remaining primary studies. Subsequently, he summarized the main ideas stated in several sentences in the selected study. He continued to encode the summarized ideas (from each selected study) to generate codes, concepts, and categories and subcategories. For example, when answering RQ1 (mined architectural information), he referred to these sentences in the selected study {[S27]} that state: “\textit{(...) Implementation of a tactic often spreads across more than one source file. Consequently, domain topics that motivate an architectural tactic are often not fully presented in the tactical file itself but also appear in neighboring files (i.e., files which use tactical files or are used by tactical files). Therefore, we need to identify the whole context in which a tactic is implemented (...)} 
In this case, he summarized and encoded those sentences as the code “identify the whole context in which a tactic is implemented”. Afterwards, the first author grouped this code into a higher-level concept (i.e., “tactic and context”). Then, he applied constant comparison to compare the concepts identified in one summarized idea with the concepts that emerged from other summarized ideas to identify the concepts which have similar semantic meanings. He proceeded to group similar concepts into main categories and subcategories. For example, the concept “tactic and context” was merged into the subcategory “architectural tactic and context relationship” that was further categorized into the category “design relationship” (see Table \ref{minedArchitecturalInfo}). It is worth noting that some primary studies (e.g., [S47]) may not explicitly clarify the concrete categories of mined architectural information. For such cases, if a study claims that it mines architectural information without providing a concrete type of this information, we thoroughly checked the content of that study including examples or excerpts from software repositories (if any) that show the mined architectural artifacts, as well as, the dataset of the study in the replication package (if available). Doing so, we tried to examine and categorize the data for identifying the concrete types of mined architectural information. 
To mitigate the personal bias during the formal data analysis, the second and third authors of this SMS reviewed and validated the generated codes, concepts, categories, and subcategories. The results of this SMS are provided in Section \ref{Results}.
\section {Results} \label{Results}

\subsection{Overview of the Selected Studies}\label{StudiesDemographic}
In this section, we present an overview of the selected primary studies from two perspectives (i.e., publication years and venues).

\textbf{(a) Distribution of the selected studies over the years} 
 
\begin{figure} [!h]
  \centering
   \includegraphics[width=0.8\linewidth]{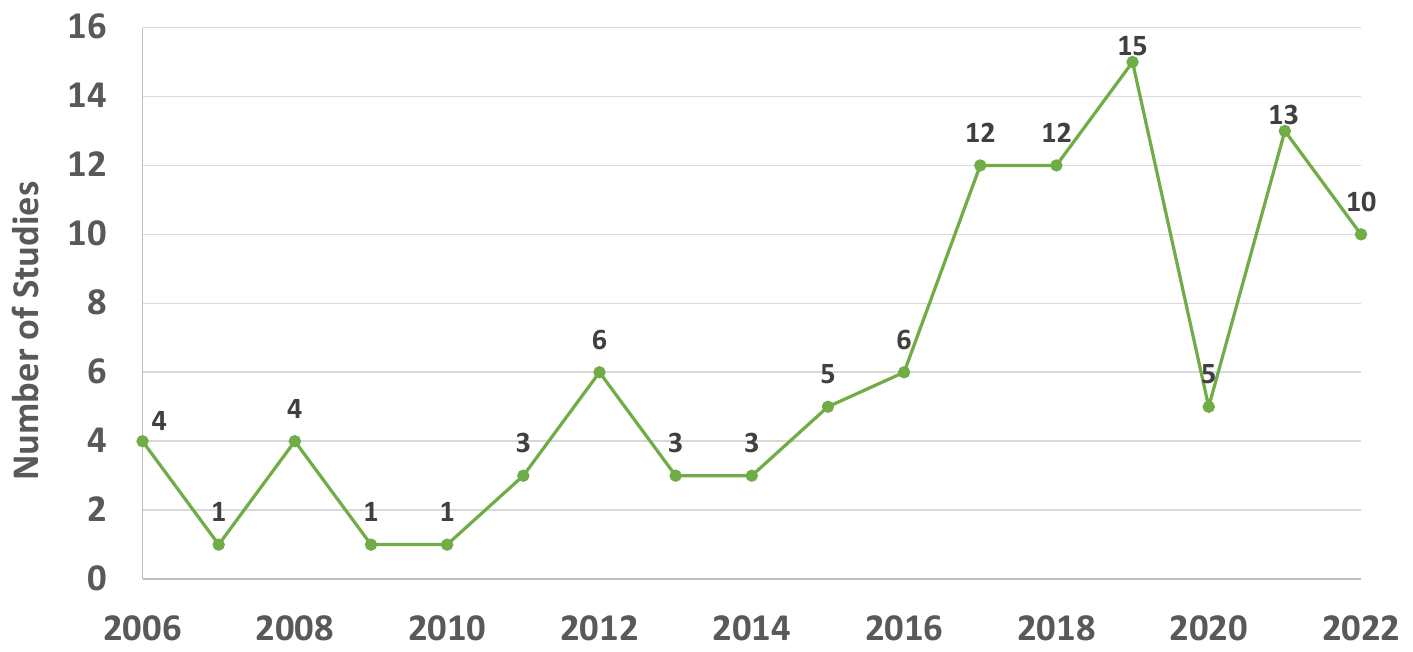}
 \caption{Distribution of the selected studies over the years}
 \label{YearsAndStudies}
 \end{figure}

To better understand the publication years of the selected studies, we summarized the publication number of these studies per year (from 2006 till 2022) and analyzed the growth of publication number as time goes by. 
Figure \ref{YearsAndStudies} shows the distribution of the 104 selected studies over the past sixteen years (i.e., 2006–2022). It is clear that there are very few studies (i.e., 11 studies) published from 2006 to 2010 on mining architectural information. However, we can see a steady amount of attention on mining architectural information from 2011 to 2022 (2-13 studies were published per year). Overall, we found evidence of a fair increasing interest in the software architecture community on mining architectural information.

 \begin{figure} [!h]
 \centering
  \includegraphics [scale=0.35]{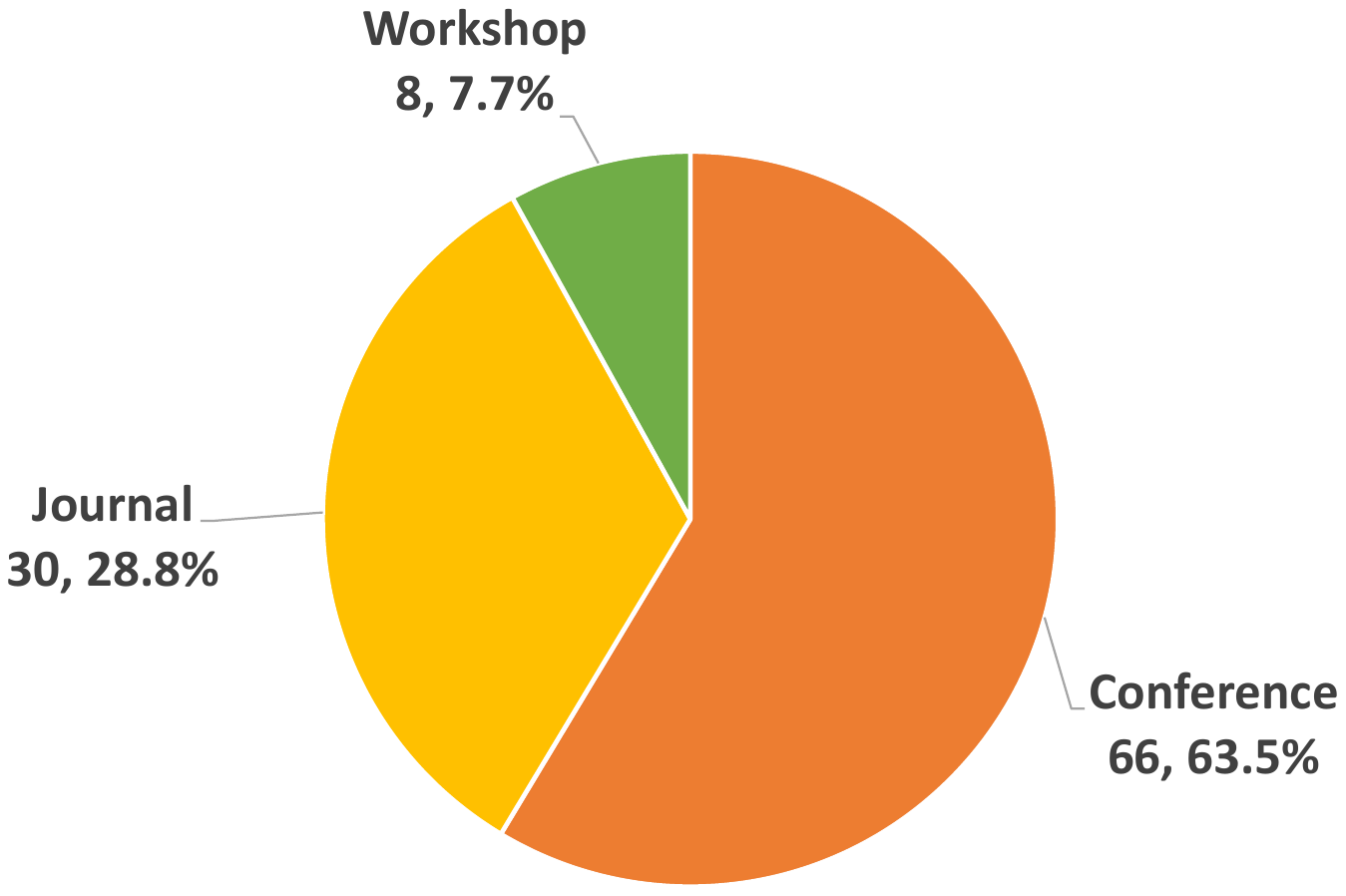}
 \caption{Publication venues based on the distribution of the selected studies}
 \label{StudiesAndVenues}
\end{figure}

  
   
 
\textbf{(b) Distribution of the selected studies in publication venues}
 
The selected studies were published in 55 venues. We only list the venues that publish more than one selected studies (see the Publication Venues in the Supplementary Material \citep{dataset}). Note that, merged conferences were only counted as one conference, for example, Working International Conference on Software Architecture (WICSA) and International Conference on Software Architecture (ICSA) were counted as one conference. As shown in the Publication Venues in \cite{dataset}, some venues tend to publish more studies on mining architectural information. For instance, the dedicated conferences on software architecture record the most publications (i.e., 17), which are International Conference on Software Architecture (ICSA) and European Conference on Software Architecture (ECSA) with 10 and 7 publications, respectively. The conference with the most number of studies (i.e., 10 studies) is ICSA. In terms of journals, both the Journal of Systems and Software and IEEE Transactions on Software Engineering index the most studies with 6 and 5 studies, respectively. As shown in Figure \ref{StudiesAndVenues}, the 104 studies cover three types of publication venues, i.e., conference, journal, and workshop. Most of the selected studies were published in conferences (i.e., 63.5\%, 66 out of 104) compared with journals (i.e., 28.8\%, 30 out of 104) and workshops (i.e., 7.7\%, 8 out of 104), which indicates that conferences are the popular venues to disseminate the work on mining architectural information.

\subsection{Results of RQ1: Mined Architectural Information}\label{ResultsOfRQ1}
As described in Section \ref{DataSynthesis}, a hybrid approach (which is a combination of open coding \& constant comparison, predefined classifications, and descriptive statistics) was used to analyze the extracted data item D4 (see Table \ref{DataExtraction}) and identify the mined architectural information from software repositories. Our data synthesis generated 7 main categories and 29 subcategories of mined architectural information (see Table \ref{minedArchitecturalInfo}), in which \textit{architectural description} (69.2\%, 72 out of 104 studies), \textit{architectural solution} (42.3\%, 44 out of 104 studies), \textit{architectural decision} (30.8\%, 32 out of 104 studies), \textit{system requirement} (29.8\%, 31 out of 104 studies) are the top three most mined architectural information to support the development process. Note that, one study (e.g., [S1]) may utilize an approach to mine more than one type of architectural information and this affects the accuracy of the statistical representation (i.e., the total percentage is more than 100\% in Table \ref{minedArchitecturalInfo}).

\textbf{Architectural description} is one of the important sources used for communicating and sharing architectural information about the high-level design of a software system \citep{clements2003documenting}. For example, the information in architectural description, such as architectural views, which describe systems in multiple views (e.g., logical view, development view), enable the architecture to be communicated and understood by various stakeholders (e.g., architects, developers, and project managers) \citep{469759}. 
69.2\% of the selected studies (72 out of 104) proposed approaches and tools for mining or recovering architectural descriptions of software systems from several artifacts (e.g., code {[S29]}{[S39]}{[S46]}{[S51]}) that are contained in various software repositories, such as VCS (e.g., GitHub), to assist the development process. We further classified the mined architectural information in the \textit{architectural description} category by using the conceptual model for architectural description in the ISO 42010:2011 standard \citep{6129467}, which defines a set of architectural elements that make up the architectural description model \citep{6129467}. 
As shown in Table \ref{minedArchitecturalInfo}, we collected five subcategories of mined architectural information (e.g., architectural model, architectural view) in the \textit{architectural description} category. 

\textit{Architectural model} is a partial abstraction of a system. Architectural model is normally a diagram that can be created by using specific notations, such as Unified Model Language (UML), in which the primary aim is to illustrate a specific set of architectural design elements (e.g., components) inherent in the structure and design of a system. 
The architecture modeling process can be bottom-up, by which details of the system are built utilizing knowledge about components and interconnections and how they compose together to realize the characteristics of the system. Alternatively, it can be top-down, by which details of the components and interconnections are extracted from knowledge of the whole system. 
Architectural model is the most mined (20 out of 72 studies) architectural information in architectural description category. For example, Granchelli \textit{et al}. {[S39]} developed a semi-automatic approach and tool (i.e., MicroART) based on Model-Driven Engineering (MDE) and Domain-Specific Language (DSL) for automatically recovering architectural models of microservice-based systems from source code. The proposed approach can also allow an architect to manually refine the initial extracted architecture model into refined architectural models that are suitable for his/her needs, such as for performing architecture change impact analysis. 

\begin{figure}
  \centering
   \subfloat[Architectural view (i.e., development view) mined from a software system ({[S81]})]{\includegraphics[width=.5\linewidth]{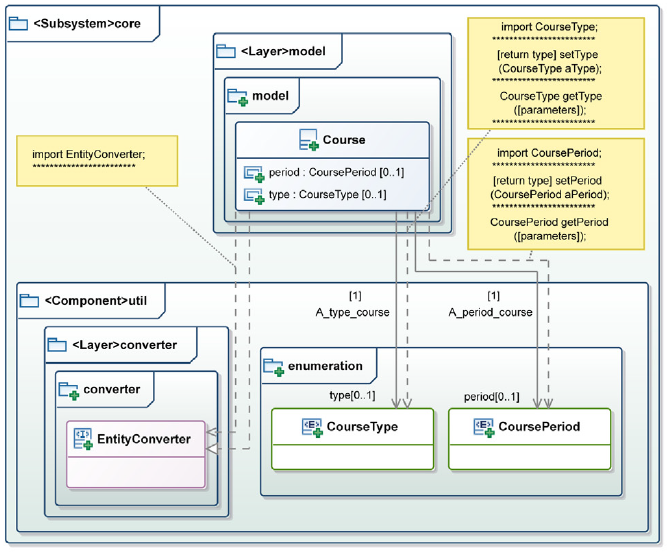}}\hfill
   \subfloat[Architectural rationale mined from the Stack Overflow site ({[S4]})] {\includegraphics[width=.45\linewidth]{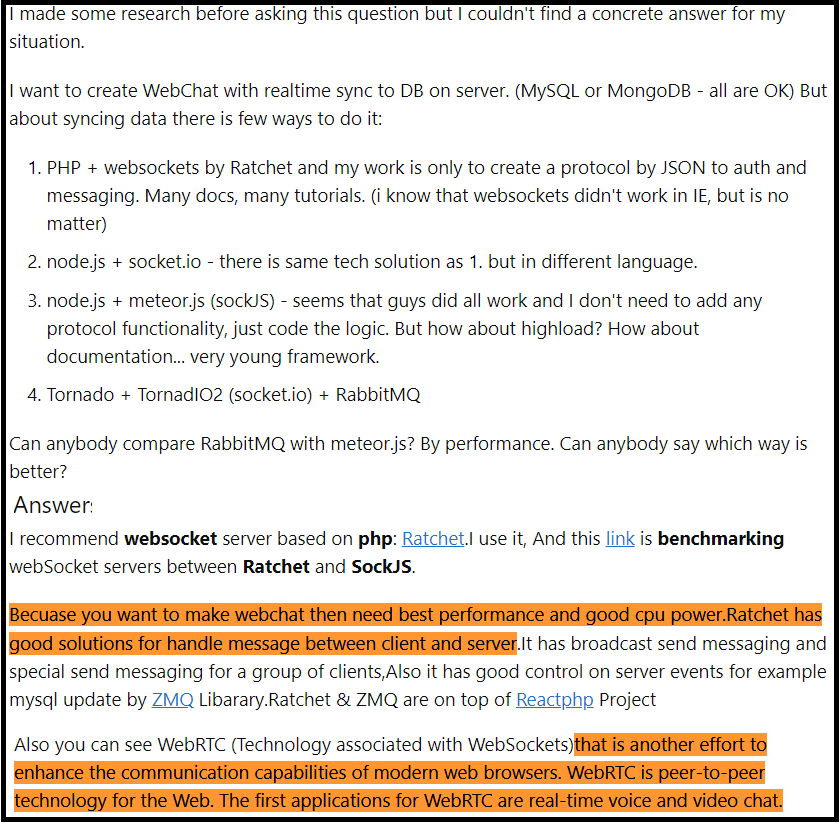}} \hfill 
   \caption{Examples of mined architectural information in the architectural description category}\color{black}
   \label{ExamplesArchitecturalDescriptionElements}
\end{figure}

\textit{Architectural view} represents the system from the viewpoint of different stakeholders, such as developers, system analysts, project managers, and end-users. The architect chooses and develops a set of views that will enable the architecture to be communicated to, and understood by, the stakeholders of the system. In capturing or representing the system in architectural views, the architect will typically create one or more architecture models (in the form of diagrams), possibly using specific notations (e.g., UML). An architecture view comprises selected parts of the system, chosen to demonstrate to a particular stakeholder or group of stakeholders that their concerns are being adequately addressed in the architecture. 
18 out of 104 studies proposed approaches and tools for mining architectural views to support various architecture design tasks. For example, in {[S46]} an architecture recovery approach and a prototype tool (named ARCADE) were developed to help software engineers (e.g., architects) automatically extract and recover various architectural views (including development views) from source code. The mined architectural views can assist in determining, for example, when, how, and to what extent system-level or component-level architectural views evolve during software development. Figure \ref{ExamplesArchitecturalDescriptionElements}(a) depicts an example of mined architectural view (i.e., development view) from a software system.

\textit{Architectural rationale} records explanation, justification, and reasoning about architectural design decisions that are made in software architecture design. 
Lopez et al. {[S24]} presented an ontology-based approach named TREx (Toeska Rationale Extraction) for extracting, representing, and exploring architectural rationale information from text documents (such as email, wiki, and meeting notes). Specifically, this approach consists of three components: (1) pattern-based information extraction to recover rationale, (2) ontology-based representation of rationale and architectural concepts, and (3) facet-based interactive exploration of rationale. Initial results from applying TREx suggest that architectural rationale (such as reasons behind architectural decisions) can be semi-automatically extracted and mined from a project’s unstructured text documents. The mined architectural rationale provides several benefits in software development, such as rationale reuse, rationale auditing, and architect training. Figure \ref{ExamplesArchitecturalDescriptionElements}(b) shows an example of architectural rationale mined from a Q\&A site (i.e., Stack Overflow). 

\textit{System of interest} denotes a specific type of a system (such as a microservice system) whose architecture is under exploration during the development. Some studies proposed approaches and tools for mining architectural information from specific types of systems. For example, Soldani et al. in [S29] proposed an approach and toolchain (i.e., $\mu$TOSCA) for mining and refactoring architectural smells in microservice systems.

\textit{Architectural concerns} pertain to any aspect of the system's functioning, development, or operation, including considerations such as performance, reliability, security, distribution, and availability. For example, in {[S66]}, Gokyer et al. developed NLP and ML-based techniques to automatically mine architectural concerns from non-functional requirements expressed in plain text. The mined architectural information (i.e., architectural concerns) can guide architects in making design decisions effectively. \newline

\textbf{Architectural solutions} are the fundamental building blocks in architecture design, and they are used to address architecture design concerns \citep{SA2012}. 44 out of 104 studies (42.3\%) researched on mining architectural solutions to support the development. We identified three subcategories of architectural solutions that are mined from repositories, namely architectural pattern, architectural tactic, and framework (see Table \ref{minedArchitecturalInfo}). For example, Liu et al. {[S63]} utilized a semi-automatic approach for mining \textit{architectural pattern} use scenarios and related pattern pairs from specific software development blogs and Q\&A site (i.e., Stack Overflow). This mined architectural information can aid developers to acquaint with the usage and relatedness of architecture patterns in today’s modern software development. 
On the other hand, Chinnappan et al. {[S9]} employed a manual approach (i.e., qualitative data analysis) to mine energy-aware \textit{architectural tactics} (such as energy savings mode tactic, stop current task \& recharge tactic) for Robotic Operating System (ROS) based software systems from GitHub, ROS Answers, and Stack Overflow. To do so, they carried out a multi-phase study that resulted in seven energy-awareness architectural tactics. To foster the applicability of the identified tactics even beyond the ROS community, they described these tactics in a generic implementation-independent manner by utilizing diagrams inspired by UML component and sequence diagram notations. These energy-aware architectural tactics can serve as guidance for roboticists, as well as other developers interested in architecting and implementing energy-aware software. Then again, M\'{a}rquez et al. {[S26]} proposed a semi-automatic approach for mining \textit{frameworks} in OSS for addressing scalability concerns in microservice applications, and these frameworks can assist developers in designing scalable microservice applications.

\textbf{Architectural decisions}: Mining and understanding architectural decisions can help inform future architectural decisions and implementation choices, and can avoid introducing architectural inefficiencies \citep{Shahbazian2018RecoveringAD} or architectural decay later \citep{Shahbazian2018RecoveringAD}. Thus, 32 out of 104 studies (30.8\%) focused on mining architectural decisions to assist the development. We categorized the mined architectural decisions into four subcategories (including structural decision, behavioral decision, and technology decision) by following existing classifications of architectural design decisions in \citep{kruchten2004ontology} and \citep{SA2012} (see Table \ref{minedArchitecturalInfo}). For example, Bhat \textit{et al}. {[S22]} employed an ML-based approach to mine and classify architectural decisions from issue tracking systems (e.g., Jira) into three subcategories, namely \textit{structural}, \textit{behavioral}, and \textit{non-existence decisions}. The mined architectural decisions in this category can help architects and developers utilize these decisions in similar development contexts. Figure 6(a) illustrates an example of mined architectural decisions (i.e., structure decisions) from an issue tracking system.

\textbf{System requirements}: Analyzing requirements to relate and map them to their corresponding architectural elements (e.g., architectural components) has become one of the major challenges faced by architects during the development process \citep{casamayor2012functional}. The failure of a high percentage of software projects is often caused by, for example, the lack of proper requirements analysis, incomplete requirement specifications, and changing requirements, among others~\citep{hull2005requirements}. 31 out of 104 studies (29.8\%) focus on mining the requirements of systems from repositories to assist architecting process (see Table \ref{minedArchitecturalInfo}).
We further categorized the mined system requirements into three subcategories: Quality Attributes (QAs), functional requirement, and constraint by following the classification described in \citep{cervantes2016designing}. 
For instance, in {[S66]}, Gokyer et al. used an approach based on NLP and ML techniques for automatically mining \textit{Non-Functional Requirements (NFRs)} expressed in plain text and mapping these NFRs to architectural elements, such as architectural components, in the problem domain. The mined architectural information can guide architects in making architectural decisions effectively. 
On the other hand, in {[S43]}, Casamayor et al. presented an approach based on NLP and ML techniques to semi-automatically mine and analyze \textit{functional requirements} (from the textual description of requirements) that will become the responsibilities of certain architectural components in the system, in order to help bridge the gap between requirements analysis and architectural design. Then again Soliman et al. {[S3]} developed a Web-based search engine for searching several types of architectural information (including \textit{constraints}) from Q\&A sites. 

 

\textbf{Architectural technical debt}: Architectural Technical Debt (ATD) is incurred by architecture decisions that consciously or unconsciously compromise system-wide QAs, particularly maintainability and evolvability \citep{li2014architectural}. Typical ATD includes violations of architecture design principles or rules \citep{li2015systematic}. ATD needs to be identified and removed since it is harmful to the system’s long-term health \citep{li2015systematic}. 24 out of 104 studies (23.0\%) developed approaches for detecting and mining ATD for different purposes, such as mining ATD for further management in order to keep the accumulated ATD under control {[S37]}. We categorized the mined architectural information related to ATD into three subcategories, namely architectural smell (e.g., Unstable Interface, Modularity Violations, Hublike Dependency, Chatty Service, Tiny Service), architectural anti-pattern (e.g., Clique, Crossing), and architectural compliance issue (see Table \ref{minedArchitecturalInfo}). For instance, in {[S71]}, D\'{ı}az-Pace et al. developed an approach based on Link Prediction (LP) techniques and an ML classification model to mine two types of \textit{architectural smells}, namely cyclic dependency and hub-like dependency from OSS projects. On the other hand, Mo et al. {[S64]} presented an approach for automatically detecting and mining six types of \textit{architecture anti-patterns} (including unstable interface, modularity violation groups), defined as connections among files that violate design principles. 
This mined architectural information can be used by architects to pinpoint architectural anti-patterns in the systems, quantify their severity, and determine priorities for refactoring. Maffort et al. {[S34]} proposed an approach that relies on four heuristic models for detecting and mining \textit{architectural compliance issues}. Specifically, the approach mines the absences (i.e., something expected is not found) and divergences (i.e., something prohibited is found) in source code based architectures. The mined architectural compliance issues can be used to rapidly raise architectural deviation warnings, without deeply involving architects.

\textbf{Architectural changes}: Software changes are inevitable. There are many reasons for software changes, such as repairing defects and evolving user requirements \citep{williams2010characterizing}. When changes affect the architecture, software engineers need a comprehensive understanding of the causes of architecture changes and their impact \citep{williams2010characterizing}. This understanding is important because, as changes may affect many aspects of a system and introduce complexities in architecture, which will likely introduce a significant number of architectural issues. 18 out of 104 studies (17.3\%) analyzed architectural change information from different perspectives (e.g., classification of architectural changes {[S40]}) to support the development. 
We adopted a predefined classification of architectural changes introduced by Williams and Carver \citep{williams2010characterizing} to categorize the mined architectural change into four subcategories: \textit{perfective} — indicates new requirements and improved functionality, \textit{corrective} — addresses flaws, \textit{adaptive} — occurs for new environment or for imposing new policies, and \textit{preventative} — indicates restructuring or redesigning the system. 
For example, Mondal {[S40]} utilized an ML-based approach for mining information related to architectural changes from source code and commit messages, and labeled them into four subcategories, namely \textit{perfective change}, \textit{corrective change}, \textit{adaptive change}, and \textit{preventive change}. This mined architectural information can help developers better analyze and characterize the causes and impact of architectural changes prior to their implementations. Figure 6(b) shows an example of mined architectural change (i.e., perfective change) from the developer mailing list.

 \begin{figure}[t]
  \centering
   \subfloat[Architecture decision (i.e., structural decision) in an issue tracking system ({[S47]})]{\includegraphics[width=.5\linewidth]{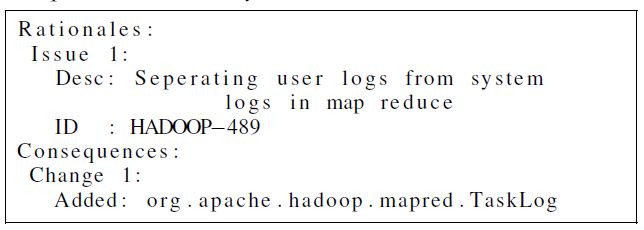}}\hfill
  \subfloat[Architecture change (i.e., perfective change) in a developer mailing list ({[S11]})] {\includegraphics[width=.45\linewidth]{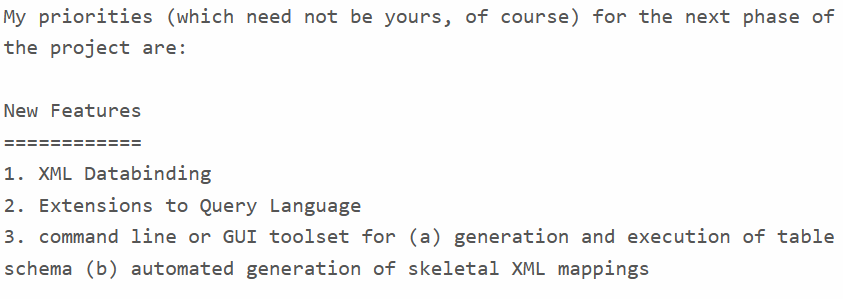}} \hfill 
   \caption{Examples mined architectural decisions and architectural changes}
 \end{figure}


\textbf{Design relationships}: An architecture of a system is designed with the aid of architectural elements (e.g., tactics and components). However, there are complex interdependency relationships between these elements, and many of the tasks faced by developers today are to deal with efficient and effective use of these elements together during the application design to meet both functional and non-functional goals~\citep{bi2018architecture}. For example, designing with a certain architecture pattern to address specific quality attribute requirements cannot be considered in isolation, but there might be other key architectural elements to be considered, such as design contexts (e.g., application context) \citep{bedjeti2017modeling}. Therefore, 17 out of 104 (16.3\%) studies researched on mining design relationships from various sources to aid the development. We categorized the mined architectural information in the Design relationship category into seven subcategories, such as Component-Component relationship, Architectural tactic-Context relationship (see Table~\ref{minedArchitecturalInfo}). 
For instance, Karthik et al. {[S20]} presented an automatic ML-based approach for mining the relationships between architectural components from the unstructured text of Q\&A site posts. Moreover, the mined \textit{Component-Component relationships} were categorized into three types, namely incompatibility, required, and recommended. This mined architectural information can help developers who work with component-based systems to effectively understand the Component-Component relationships of certain applications similar to the applications discussed on Q\&A sites. Whilst applying Architectural Tactics (ATs) to address QAs is well explored in existing works, e.g., \citep{mirakhorli2013domain, bi2021mining}, there are no guidelines for architects, who look for information on what considerations (e.g., design contexts) they need to consider when applying ATs to address QA concerns. In order to provide architects with such information, Gopalakrishnan et al. {[S27]} proposed an ML and text mining-based approach for extracting and mining typical design contexts in the source code in which architectural tactics are implemented, and analyzed the relationship between the design contexts and architectural tactics implemented in the code (i.e., \textit{Architectural tactic-Context relationship}).
On the other hand, Wijerathna \textit{et al.} {[S87]} employed an automatic ML-based approach for mining design contexts such as platform context (e.g., software context) from Stack Overflow posts, and manually structured the design relationships between design contexts and architectural patterns used in practice (i.e., \textit{Architectural pattern-Context relationship}). This information can assist software engineers (e.g., architects) in choosing the appropriate architectural patterns with respect to design context. In {[S31]}, Velasco-Elizondo et al. used an approach based on an information extraction technique (i.e., entity extraction) and knowledge representation (i.e., ontology) to automatically analyze and mine architecture patterns considering specific QAs (i.e., \textit{Architectural pattern-Quality attribute relationship}) from architectural pattern descriptions. The mined architectural information in this subcategory can facilitate developers to select architectural patterns through knowing whether specific QAs are promoted or inhibited. 

\begin{landscape}
\small
\begin{longtable}{p{10em}p{19em}p{19em}p{2.5em}}
\caption{Categories and subcategories of mined architectural information} \label{minedArchitecturalInfo} \\\hline
\textbf{Category} & \textbf{Subcategory} & \textbf{Studies} & \textbf{Count} \\\hline  
\endfirsthead
\multicolumn{4}{c}
{{\bfseries }}\\
\endhead 
\multicolumn{4}{r}{{}} \\ 
\endfoot
\hline
\endlastfoot
{Architectural description (69.2\%, 72 out of 104)} 
                                   
                                         & Architectural model 
                                         & {[S11]} {[S12]} {[S29]} {[S30]} {[S36]} {[S37]} {[S39]} {[S46]} {[S49]} {[S51]} {[S54]} {[S62]} {[S64]} {[S67]} {[S68]} {[S73]} {[S75]} {[S80]} {[S83]} {[S86]} 
                                         & 20 \\ \cline{2-4}

                                         & Architectural view
                                         & {[S8]} {[S16]} {[S18]} {[S30]} {[S35]} {[S37]} {[S39]} {[S46]} {[S47]} {[S54]} {[S59]} {[S61]} {[S64]} {[S68]} {[S98]} {[S70]} {[S81]} {[S102]}   
                                         & {18} \\\cline{2-4}
                                        
                                         & Architectural rationale 
                                         & {[S1]} {[S2]} {[S4]} {[S11]} {[S18]} {[S23]} {[S24]} {[S25]} {[S26]} {[S28]} {[S38]} {[S47]} {[S79]}
                                         & {13} \\\cline{2-4}

                                         & System of interest
                                         & {[S11]} {[S26]} {[S29]} {[S39]} {[S60]} {[S72]} {[S85]} {[S100]} {[S103]}
                                         & {9} \\\cline{2-4}
                                        
                                         & Architectural concern
                                         & {[S4]} {[S6]} {[S7]} {[S8]} {[S9]} {[S11]} {[S38]} {[S66]} 
                                         & {8} \\\cline{1-4}

{Architectural solution (42.3\%, 44 out of 104)}
                                        & Architectural pattern 
                                        &{[S1]} {[S2]} {[S3]} {[S4]} {[S16]} {[S23]} {[S24]} {[S28]} {[S32]} {[S44]} {[S49]} {[S50]} {[S52]} {[S53]} {[S54]} {[S61]} {[S63]} {[S69]} {[S70]} {[S82]} {[S79]}   
                                        & {21} \\ \cline{2-4}
                                        
                                        & Architectural tactic 
                                        & {[S1]} {[S4]} {[S6]} {[S7]} {[S13]} {[S26]} {[S9]} {[S24]} {[S49]} {[S50]} {[S53]} {[S56]} {[S57]} {[S79]} 
                                        & {14} \\ \cline{2-4}
                                         
                                        & Framework 
                                        & {[S3]} {[S4]} {[S14]} {[S16]} {[S26]} {[S49]} {[S50]} {[S52]} {[S61]}  
                                        & {9} \\ \cline{1-4}  
                                    
{Architectural decision (30.8\%, 32 out of 104)}
                                        & Technology decision  
                                        & {[S2]} {[S3]} {[S4]} {[S14]} {[S15]} {[S16]} {[S18]} {[S23]} {[S33]} {[S38]} {[S53]} {[S85]}
                                        & {12} \\ \cline{2-4}
                                        
                                        & Structural decision 
                                        & {[S17]} {[S18]} {[S19]} {[S22]} {[S23]} {[S32]} {[S33]} {[S47]} {[S54]} {[S85]}  
                                        & {10}\\ \cline{2-4}
                                    
                                        & Behavioral decision  
                                        & {[S18]} {[S19]} {[S22]} {[S33]} {[S42]} {[S47]} {[S54]}   
                                        &  {7} \\ \cline{2-4} 
                                        
                                        & Ban or non-existence decision  
                                        & {[S19]} {[S22]} {[S33]}    
                                        &  {3} \\ \cline{1-4}  
                                        
{System requirement (29.8\%, 31 out of 104)}   
                                        & Quality attribute  
                                        & {[S1]} {[S2]} {[S3]} {[S4]} {[S8]} {[S11]} {[S16]} {[S17]} {[S23]} {[S24]} {[S26]} {[S28]} {[S32]} {[S42]} {[S61]} {[S66]} {[S76]} {[S83]}  
                                        & {18} \\ \cline{2-4}
                                        
                                        & Functional requirement 
                                        & {[S1]} {[S4]} {[S17]} {[S43]} {[S45]} {[S54]} {[S55]} {[S83]} {[S79]}    
                                        & {9} \\ \cline{2-4}

                                        & Constraint 
                                        & {[S1]} {[S2]} {[S3]} {[S54]}     
                                        & {4} \\ \cline{1-4} 

{Architectural technical debt (23.0\%, 24 out of 104)}  
                                        & Architectural smell
                                        & {[S5]} {[S29]} {[S65]} {[S71]} {[S84]} {[S88]} {[S90]} {[S92]} {[S93]} {[S94]} {[S97]} {[S98]} {[S99]} {[S100]}
                                        & {14} \\ \cline{2-4} 

                                        & Architectural anti-pattern
                                        & {[S37]} {[S60]} {[S64]} {[S72]} {[S95]} {[S104]} 
                                        &  {6} \\ \cline{2-4}

                                        & Architectural compliance issue
                                        & {[S30]} {[S34]} {[S36]} {[S73]} 
                                        &  {4}  \\ \cline{1-4}                                        

{Architectural change (17.3\%, 18 out of 104)} 
                                        & Perfective change 
                                        & {[S17]} {[S18]} {[S32]} {[S35]} {[S40]} {[S11]} {[S77]} 
                                        & {7} \\ \cline{2-4}
                                        
                                        & Corrective change
                                        & {[S35]} {[S32]} {[S40]} {[S46]} {[S77]} 
                                        & {5} \\ \cline{2-4}
                                        
                                        & Adaptive change
                                        & {[S35]} {[S40]} {[S77]} 
                                        & {3} \\ \cline{2-4}
                                        
                                        & Preventative change 
                                        & {[S35]} {[S40]} {[S77]}
                                        & {3} \\ \cline{1-4}                                        

{Design relationship (16.3\%, 17 out of 104)} 
                                        & Component-Component relationship
                                        & {[S12]} {[S20]} {[S48]} {[S59]} {[S67]} {[S69]} {[S74]} {[S101]} {[S103]}
                                        & {9} \\ \cline{2-4}
                                        
                                        & Architectural tactic-Context relationship
                                        & {[S6]} {[S27]}   
                                        &   {2} \\ \cline{2-4}
                                        
                                        & Architectural pattern-Architectural pattern relationship 
                                        & {[S63]} {[S78]}
                                        &  {2} \\ \cline{2-4}

                                        & Architectural pattern-Context relationship
                                        & {[S87]}
                                        &  {1} \\ \cline{2-4}
                                        
                                        & Architectural pattern-Quality attribute relationship 
                                        & {[S31]}   
                                        &  {1} \\ \cline{2-4}
                                        
                                        & Architectural tactic-Quality attribute relationship
                                        & {[S6]}    
                                        &  {1} \\ \cline{2-4}
                                        
                                        & Design pattern-Architectural tactic relationship
                                        & {[S21]}   
                                        &  {1} \\ \cline{1-4}                                        
\end{longtable}
\end{landscape}
\normalsize


 \begin{tcolorbox}[colback=gray!5!white,colframe=gray!75!black,title=Key Findings of RQ1]
 \textbf{Finding 1}: 
 A broad range of architectural information has been mined to support the development. The mined architectural information, such as \textit{Architectural tactic-Context relationship}, can assist software engineers (e.g., developers) in choosing and implementing the appropriate architectural tactics with respect to design context. 

 \end{tcolorbox}

\subsection{Results of RQ2: Sources used for Mining Architectural Information}\label{ResultsOfRQ2}
\begin{figure}
 \centering
  \includegraphics[width=.9\linewidth]{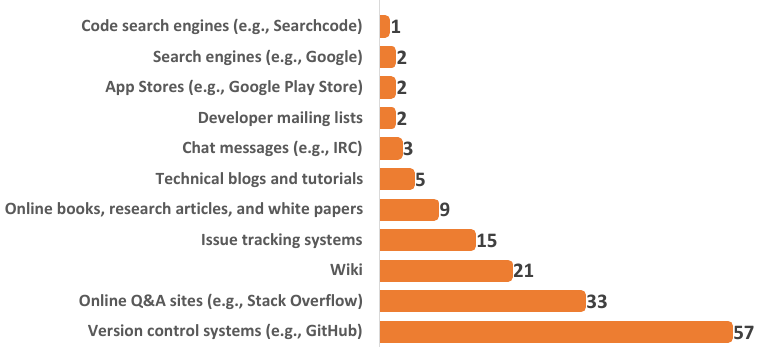}
 \caption{Sources used for mining architectural information}
 \label{SourceMinedFigure}
\end{figure}

We applied open coding \& constant comparison and descriptive statistics to analyze the extracted data item D5 and D6 (see Table \ref{DataExtraction}) for answering RQ2. We first categorized the reported sources into eleven core categories (see Figure \ref{SourceMinedFigure}). Note that, GitHub describes itself as “\textit{an Internet hosting service for software development and version control using Git}”, and in this SMS we opted to encode it as a Version Control System (VCS). As shown in Figure \ref{SourceMinedFigure}, the most frequently used source when mining architectural information is \textit{VCS}, e.g., GitHub, (used by 57 studies), followed by \textit{Online Q\&A sites}, e.g., Stack Overflow, (used by 33 studies), and \textit{Wiki} (used by 21 studies). In the following paragraph, we provide examples for the five most employed sources in mining architecture information. 

Ghorbani et al. {[S30]} collected a set of open-source Java applications from \textit{VCS} (specifically, GitHub) and mined architectural inconsistencies from those applications. Soliman et al. {[S4]} collected architecture related posts in \textit{Online Q\&A sites} (specifically, Stack Overflow) and mined architectural knowledge for technology decisions. Gilson et al. {[S14]} gathered data from two sources (i.e., \textit{Wiki} and \textit{Online Q\&A sites} (specifically, Stack Exchange)) for mining alternative architectural solutions (such as architectural pattern) to a given architectural problem, in order to provide assistance to the decision-making architect. Shahbazian et al. {[S35]} crawled several open source projects from \textit{Issue tracking systems} (specifically, Jira) to detect and mine implementation issues that lead to possibly unintentional architectural design decisions and subsequently architectural changes, such as perfective change \citep{williams2010characterizing}. In {[S78]}, Kamal and Avgeriou proposed a manual approach for mining architectural pattern-architectural pattern relationships from books, research articles, and white papers. Note that, since some studies used more than one source, the sum of the sources reported in Figure \ref{SourceMinedFigure} exceeds the total number of the selected studies (i.e., 104 studies).
 
\begin{figure} 
 \centering
  \includegraphics [scale=0.4]{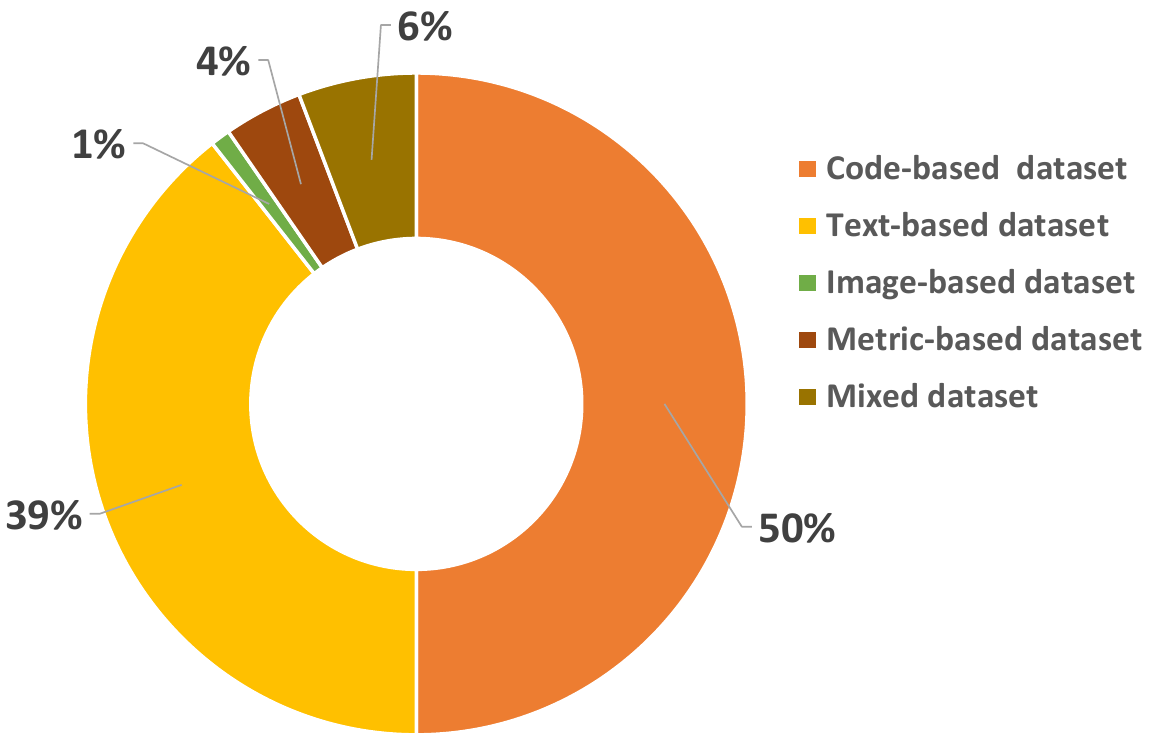}
 \caption{Datasets used for mining architectural information}
 \label{DatasetMinedFigure}
\end{figure}

Moreover, the sources used in the primary studies contain various dataset types. It is essential to analyze dataset types since the relationship between the type of implicit feature being extracted and the approach (specifically, for semi-automatic and automatic approaches) has a dominating influence on approach selection. Besides, the difference in dataset types used in the primary studies also impacts the data pre-processing technique adoption for semi-automatic and automatic approaches. Therefore, we summarized and categorized dataset types in the primary studies into five categories: \textit{code-based}, \textit{text-based}, \textit{metric-based}, \textit{image-based}, and \textit{mixed} dataset (see Figure \ref{DatasetMinedFigure}). We observe that most of the primary studies adopted \textit{code-based datasets} to conduct their experiments, where 50\% (52 out of 104) of the studies adopted code-based datasets for mining architectural information to support several important architecting activities, such as architecture recovery (e.g., {[S39]}{[S41]}{[S47]}{[S101]}), architectural conformance checking (e.g., {[S81]}{[S85]}{[S86]}), and architecture maintenance and evolution (e.g., {[S59]}{[S89]}{[S97]}{[S102]}). 
This phenomenon indicates that source code (as the most valuable data in software development) can provide more complete information (e.g., low-level and high-level information) for software systems in contrast with other non code-based datasets. \textit{Text-based dataset} is the second most utilized dataset type (i.e., 41 out 104 studies, 39\%) and we observed that text-based dataset includes the dataset types, such as architectural issues {[S1]}, requirements documentation (e.g., {[S55]}), architectural documentation (e.g.,{[S45]}), code comments (e.g., {[S73]}), email discussions (e.g., {[S23]}), and log information (e.g., {[S96]}). On the other hand, 4 (out of 104, 4\%) studies used \textit{metric-based datasets}. For example, Behnamghader et al. {[S46]} employed a metric-based dataset (e.g., Architecture-to-Architecture (a2a)) for quantifying architectural change and decay across the development history of software systems. One study (i.e., {[S62]}) used an \textit{image-based dataset} for mining architectural models to support the architecture recovery activity. In addition, a few studies used \textit{mixed datasets}, such as a combination of source code and commit messages (e.g., {[S40]}) for mining architectural information. 

\begin{tcolorbox}[colback=gray!5!white,colframe=gray!75!black,title=Key Findings of RQ2]
\textbf{Finding 2}:
The majority of primary studies utilized \textit{code-based} and \textit{text-based} datasets for mining architectural information. Moreover, these datasets were mainly collected from \textit{VCS}, \textit{Online Q\&A sites}, and \textit{Wiki}. One possible reason is that code-based and text-based datasets are usually the major datasets available in open-source repositories. Furthermore, \textit{image-based datasets} are rarely used and one reason could be that there is a lack of open source image-based datasets in software repositories.
\end{tcolorbox}

\subsection{Results of RQ3: Supported Architecting Activities}\label{ResultsOfRQ3}
As mentioned in Section~\ref{DataSynthesis}, a hybrid approach (which is a combination of open coding \& constant comparison, predefined classifications of architecting activities by \cite{hofmeister2007general}, \cite{tang2010comparative}, and \cite{li2013application}, and descriptive statistics) was used for analyzing the data item D7 to answer RQ3. \cite{hofmeister2007general} presented a general model for architecture design, which consists of three main activities: architecture analysis, architecture synthesis, and architecture evaluation. \cite{tang2010comparative} extended this general model by adding two architecting activities (i.e., architecture implementation and architecture maintenance) in the architecture life cycle. These two architecting activities emphasize that architecture is not only an important asset for design but also for the later stage in the software development life cycle. \cite{li2013application} extended the aforementioned architecting activities by adding five more activities (i.e., architecture recovery, architectural description, architecture understanding, architecture impact analysis, and architecture reuse) from the perspective of Architecture Knowledge (AK) management (e.g., AK reuse) to support the architecture life cycle. Therefore, we followed the ten predefined architecting activities when answering RQ3. However, through our qualitative data analysis (see Section \ref{DataSynthesis}), we obtained one additional architecting activity, namely “architecture conformance checking”.  

Moreover, to comprehensively understand how the identified architecting activities are supported, we analyzed the primary studies to explore how these studies contribute to architecting activities through the proposed approaches and tools (results of RQ4). To achieve this, we investigated these primary studies to present the relationships between architecting activities (e.g., architecture reuse) with respect to types of problems solved (e.g., architectural change classification), approaches and/or tools used (e.g., clustering-based approaches), mined architectural information (e.g., architectural views), dataset types (e.g., code-based datasets) as well as sources (e.g., Code search engines) used in mining architectural information. 

\begin{figure}
 \centering
  \includegraphics[width=1\linewidth]{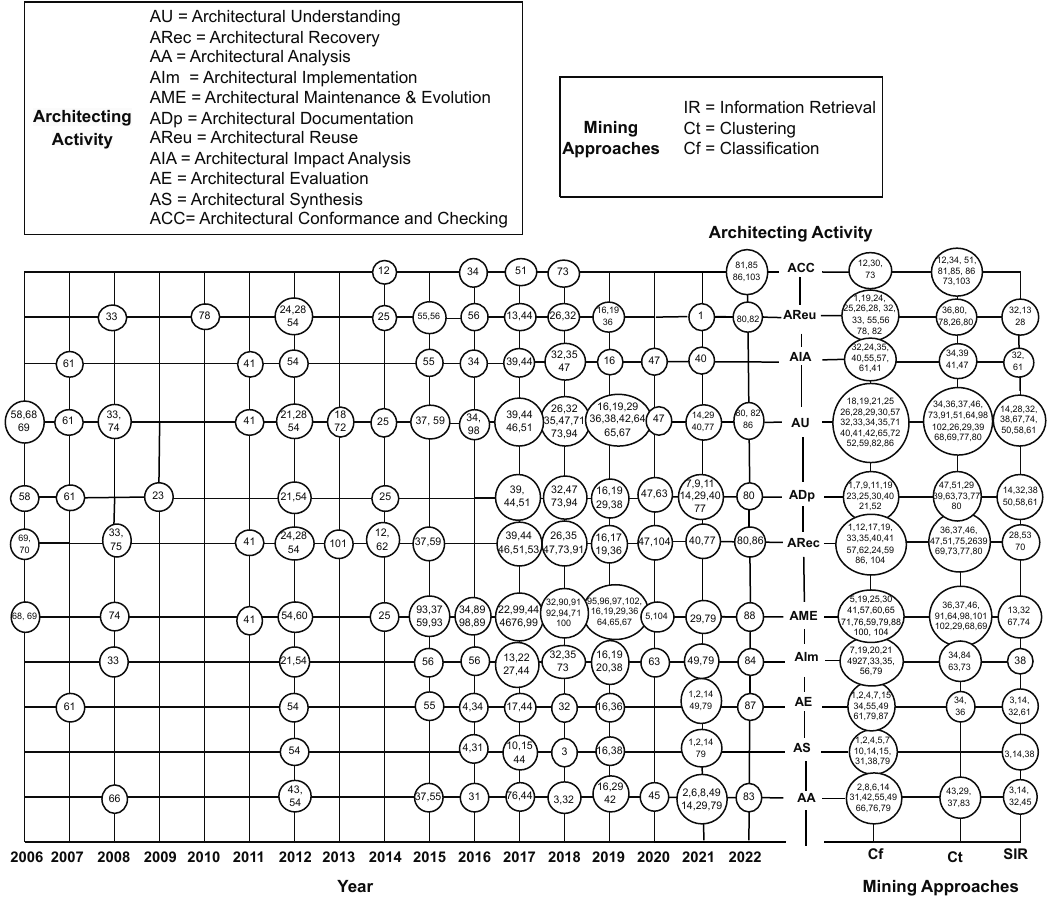}
 \caption{Bubble chart over mining approaches, architecting activities, and time period}
 \label{MiningApproachesArchActivitiesYears}
\end{figure}

We present the frequency of occurrence of the eleven architecting activities supported by the mined architectural information in the Architecting Activities sheet in the Supplementary Material \citep{dataset}. Moreover, that sheet also provides the relationships between supported architecting activities, mined architectural information, datasets, and sources used for mining architectural information as well as the relevant studies. These relationships act as a panorama to comprehensively understand the state of the research in mining architectural information. Furthermore, Figure \ref{MiningApproachesArchActivitiesYears} presents the results in a map through the studies distributed over the dimensions of architecting activities, mining approaches, and time period. In the top of Figure \ref{MiningApproachesArchActivitiesYears}, we illustrate eleven architecting activities and three categories of mining approaches. The relationships of the three elements are shown in the lower half part of the map. The left part reflects the relationship between publication year and architecting activities of the selected studies, and the number in bubbles represents the corresponding selected studies (see the Selected Studies in the Supplementary Material \citep{dataset}) in specific architecting activities. The right part shows the relationship between architecting activities and mining approaches. The number in bubbles signifies the selected studies employing a certain type of mining approach in a specific architecting activity. As shown in Figure \ref{MiningApproachesArchActivitiesYears}, most of the studies treated the problems in architecting activities as classification and clustering tasks and, therefore, these studies mostly employed classification-based and clustering-based approaches and/or tools for mining architectural information from diverse sources to support various architecting activities.

\textit{\textbf{Architecture Understanding (AU)}} is conducted to comprehend elements of an architecture design, such as architectural solutions, architectural components, and relationships between those elements (e.g., the relationships between architectural components \citep{stevanetic2014exploring}). The information gained during AU can be used as input for other architecting activities (such as architecture implementation). The AU activity is the most supported activity (i.e., 44.2\% , 46 out of 104) of the studies (see the Architecting Activity sheet in the Supplementary Material \citep{dataset}). 

\textit{Problems solved and approaches used in AU} - Various problems, such as poor understanding of architectural elements (e.g., architectural patterns), poor understating of the relationships between architectural elements are faced by software engineers during architecture design \citep{li2013application}. One common practice of mining approaches is to process textual or non-textual artifacts and mine key information to improve the understanding of those artifacts, especially for large collections of textual or non-textual artifacts. The majority of the studies in AU activity treated the problems in this activity as classification and clustering-based tasks (see Figure \ref{MiningApproachesArchActivitiesYears})). Thus, these studies mainly utilized classification-based (e.g., {[S19]}{[S21]}{[S37]}{[S40]}{[S69]}) and clustering-based (e.g., {[S67]}{[S46]}{[S74]}{[S91]}) approaches and tools for mining architectural information in order to solve different problems in this activity (see the Architecting Activities sheet in the Supplementary Material \citep{dataset}). We classified the problems solved in AU activity into seven types: understanding of architecture description (e.g., via mining architectural views {[S91]} and models {[S39]}), understanding of architectural solutions (e.g., via classification of architectural solutions {[S61]}{[S21]}), understanding of architectural decisions (e.g., via classification of architectural decisions {[S19]}), understanding of system requirements (e.g., classification of system requirements {[S83]}), understating of architectural changes (e.g., via classification of architectural changes {[S40]}{[S77]}), understanding of architectural technical debt (e.g., via detection and classification architectural smells {[S29]}{[S96]}{[S37]}), and understanding of the relationships between architectural elements (e.g., via mining Component-Component relationship (e.g., {[S67]}{[S74]}). For instance, for improving architectural description via mining architectural views, Nascimento Vale et al. {[S59]} proposed an approach called Keecle for semi-automatically identification of architecturally relevant classes that represent components from source code. The approach uses Naïve Bayes classification algorithm to classify relevant classes that provide a high-level overview of a software system in order to help developers understand the entire software system.

\textit{Datasets and sources used in AU} - Other critical problems in AU activity are the datasets and sources in which architectural information can be extracted in order to solve the identified problems. Thus, we also explored and reported datasets and sources used in AU. Four types of datasets were used in AU: code-based (e.g., {[S67]}{[S74]}), text-based (e.g., {[S19]}{[S58]}), metric-based (e.g., {[S46]}), and mixed datasets (e.g., code and documentation {[S69]}) (see the Architecting Activities sheet in the Supplementary Material \citep{dataset}). Moreover, these datasets were mainly collected from code repositories, such as GitHub (e.g., {[S47]}{[S37]}).

\textbf{\textit{Architecture Maintenance and Evolution (AME)}}: No software product is ever completely finished and all need some form of ongoing maintenance. Architectural maintenance is to adjust the architecture according to changes and faults. Architecture may evolve during its lifetime to respond to changing requirements \citep{tang2010comparative}. AME is the second most assisted architecting activity (i.e., 40.4\%, 42 out of 104 studies)  (see the Architecting Activity sheet in the Supplementary Material \citep{dataset}).

\textit{Problems solved and approaches used in AME} - Software engineers face various problems during AME, such as identification and removal of architectural technical debt (e.g., architectural smells), especially, in large-scale projects. 
The studies in AME mainly employed clustering-based and classification-based approaches to mitigate the problems in this activity. The problems solved in AME activity include detection and clustering/classification of architectural technical debt (e.g., architectural smells {[S98]}[S99][S100]), detection and classification of architectural changes (e.g., {[S46]}{[S40]}), identification and classification of architectural decisions (e.g., {[S47]}) that lead to changes in architecture. For example, for detection and clustering architectural debt, in {[S37]}, Kazman et al. presented an approach that relies on the Design Rule Space (DRSpace) analysis concept to precisely locate three types of architectural smells, namely Unstable Interface, Implicit Cross-module Dependency, and Unhealthy Inheritance Hierarchy. The approach generates a set of dependencies between the source code files of projects, clusters the source code files into DRSpace using a Design Structure Matrix (DSM), and finally identifies and visualizes the architectural smells among these files, which reduces the long-term maintenance efforts at the architecture level.

\textit{Datasets and sources used in AME} - Common datasets in AME are: code-based (e.g., {[S37]}{[S90]}{[S95]}), metric-based (e.g., {[S46]}), and mixed (e.g., code and commit messages) (e.g., {[S40]}) datasets, and these datasets were mainly collected from version control systems.

\textit{\textbf{Architecture Recovery (ARec)}} is mainly performed to recover architectural information from systems implementation \cite{schmitt2020arcade} or documentation~\citep{shahin2013recovering}. The ARec activity is the third most supported activity (i.e., 32.7\% (34 out of 104)) of the studies.  

\textit{Problems solved and approaches used in ARec} - Several problems have been solved in ARec activity including recovering architectural description (e.g., via mining architectural views {[S46]}), recovering architectural solutions (e.g., via mining architectural patterns {[S1]}), recovering architectural decisions (e.g., via extraction of architectural decisions {[S19]}), recovering system requirements (e.g., via extraction of system requirements {[S76]}) (see the Architecting Activities sheet in the Supplementary Material \citep{dataset}). 
Moreover, the studies in ARec treated the problems in this activity as classification, clustering, and Information Retrieval (IR) based tasks and used classification-based (e.g., {[S19]}{[S33]}), clustering-based (e.g., {[S41]}{[S46]}{[S47]}), and IR-based (e.g., {[S28]}{[S53]}) approaches and/or tools for mining architectural information in order to support ARec. As shown in Figure~\ref{MiningApproachesArchActivitiesYears}, clustering-based and classification-based approaches are the most used approaches in ARec activity. For example, for clustering-based approaches, Shahbazian et al. {[S47]} developed an approach named RecovAr that relies on two state-of-the-art architectural recovery techniques (i.e., Algorithm for Comprehension-Driven Clustering (ACDC) \citep{tzerpos2000accd} and Architecture Recovery using Concerns (ARC) {[S41]}) for automatically recovering architectural decisions (e.g., behavioral decisions) from source code of software projects. 

\textit{Datasets and sources in ARec} - Five types of datasets were utilized in ARec activity, namely code-based (e.g., {[S41]}{[S75]}), text-based (e.g., [S19]{[S73]}), metric-based (e.g., {[S46]}), image-based (e.g., {[S62]}), and mixed (e.g., {[S35]})) datasets. Furthermore, version control system is the common source for ARec activity.

\textbf{\textit{Architectural Description and Documentation (ADp)}}: The main goals of ADp are facilitating the expression and evolution of software systems, providing a blueprint for system development, and supporting the communication between stakeholders \citep{6129467}. The ADp activity is mentioned in 26.9\% (28 out of 104) of the studies.

\textit{Problems solved and approaches used in ADp} - Architecture is described and documented using a collection of architectural elements, such as architectural views \citep{clements2003documenting}. Problems solved in ADp activity include improving ADp via mining architectural models and views (e.g., {[S47]}{[S54]}), improving ADp via mining architectural solutions (e.g., [S1] [S23] [S50]), improving ADp via mining system requirements (e.g., {[S1]}{[S52]}). Classification-based, clustering-based, and IR-based approaches were utilized to solve the problems in ADp (see Figure \ref{MiningApproachesArchActivitiesYears}). For example, for IR-based approaches, Aman-ul-haq and Ali Babar {[S23]} proposed a tag and annotation-based knowledge identification and extraction approach that uses keyword-based search to retrieve desired architectural artifacts (such as architectural solutions) in the repository. This approach can reduce the time and effort required for searching and documenting architectural information during development. 

\textit{Datasets and sources in ADp} - Text-based (e.g., {[S11]}{[S50]}{[S52]}), code-based (e.g., {[S77]}), and mixed (i.e., code and comments {[S40]}) datasets are prevalent datasets in ADp activity, and these types of datasets are mainly collected from version control systems (e.g., GitHub), developer mailing lists, and Wiki.

\textbf{\textit{Architecture Implementation (AIm)}}: During the AIm activity, developers transform and implement architecture design into code \citep{hofmeister2007general}. The AIm activity is supported by 19.2\% (20 out of 104) of the studies (see the Architecting Activities sheet in the Supplementary Material \citep{dataset}). 

\textit{Problems solved and approaches used in AIm} - We classified the problems solved in AIm into three types: architectural solution implementation (e.g., via mining code snippets that implement architectural tactic {[S13]}), and improving architectural description via mining architectural views (development view {[S54]}), and architectural decision implementation. For example, for architectural solution implementation, in {[S13]} Mujhid et al. proposed an approach based on IR and program analysis techniques for searching and identifying which source codes from software repositories are architecturally relevant (i.e., implementing an architectural tactic). The approach uses a novel IR algorithm that automatically searches and ranks the source files implementing architectural tactics in the repositories. The approach can help developers search and use implementation examples of architectural tactics for a given technical context. 

\textit{Datasets and sources in AIm} - Frequent datasets used in AIm are: code-based (e.g., {[S13]}{[S21]}), text-based (e.g., {[S7]}{[S20]}), and mixed (e.g., code and comments {[S35]}) datasets. In addition, these datasets were collected from various sources, such as version control systems, Q\&A sites, and technical blogs and tutorials.

\textbf{\textit{Architecture Analysis (AA)}}: AA examines, filters, and reformulates architectural concerns and context that an architecture needs to address \citep{li2013application}. During AA, Architecturally Significant Requirements (ASRs) are identified from given architectural concerns and context \citep{hofmeister2007general}. The AA activity is supported by 17.3\% (18 out of 104) of the studies (see the Architecting Activities sheet in the Supplementary Material \citep{dataset}). 

\textit{Problems solved and approaches used in AA} - Classification-based approaches were predominately employed in AA to address different problems, such as identification and classification of architectural concerns, identification and classification of ASRs (see Figure \ref{MiningApproachesArchActivitiesYears}). Effective identification and classification of ASRs enables architects to perform well-focused communication with developers and users as well as prioritize the ASRs according to their importance. For instance, for classification of ASRs, in {[S42]}, Gilson et al. proposed a classification-based approach for extracting and classifying ASRs (i.e., quality attributes) from user stories in agile development. The mined architectural information guides developers to get relevant quality attributes and potential architectural key drivers during AA and provides a “bigger picture” of potential architectural drivers for early architecture decision-making. 

\textit{Datasets and sources in AA} - Common datasets used in AA is text-based (e.g., {[S8]}{[S42]}) datasets. This dataset was mainly collected from requirements documentation in Wiki, technical blogs and tutorials, and Q\&A sites.

\textbf{\textit{Architecture Reuse (AReu)}} is to reuse existing architectural elements for addressing various architectural problems, and this activity can help achieve an architecture of better quality at a lower cost~\citep{li2013application}. The AReu activity is facilitated by 17.3\% (18 out of 104) of the studies. 

\textit{Problems solved and approaches used in AReu} - Clustering-based approaches are predominately used to solve problems related to AReu (see Figure \ref{MiningApproachesArchActivitiesYears}). Among others, improving architectural solutions reuse (e.g., tactics {[S1]}{[S56]}) and improving architectural decisions reuse (e.g., {[S19]}{[S22]}) are the major problems solved in AReu. For example, in {[S1]}, Soliman et al. used a semi-automatic classification approach to mine architectural information (including architectural solutions and architectural decisions) from issue tracking systems. Among other support, the mined architectural information can support practitioners (e.g., architects) in determining architectural scenarios, in which the reuse of the architectural information (e.g., architectural solutions) from issue tracking systems could be the most suitable. 

\textit{Datasets and sources AReu} - Code-based (e.g., {[S13]}), text-based (e.g., {[S1]}{[S25]}), and mixed (e.g., code and documentation {[S69]}) datasets are the three types of datasets used in AReu activity. On the other hand, version control systems, Q\&A sites, Wiki, issue tracking systems, and technical blogs and tutorials are the prevalent sources for AReu (see the Architecting Activities sheet in the Supplementary Material \citep{dataset}).

\textbf{\textit{Architecture Evaluation (AE)}} aims to assess the architectural solutions that are proposed during AS activity against the ASRs \citep{hofmeister2007general}. During the AE activity, various factors are considered, such as pros and cons, trade-offs, and constraints of architectural solutions that result in the selection of appropriate architectural solutions for satisfying ASRs. The AE activity is supported by 16.3\% (17 out of 104) of the studies (see the Architecting Activities sheet in the Supplementary Material \citep{dataset}). 

\textit{Problems solved and approaches used in AE} - Classification-based approaches were mainly utilized to solve problems related to AE (see Figure \ref{MiningApproachesArchActivitiesYears}). Problems solved in AE activity include identification and classification of architectural solutions, architectural decisions, and system requirements. For instance, Andrzej et al. {[S14]} proposed an Architecture Decision Support System (ADSS) tool that can identify and classify various types of architectural information, such as architectural solutions (e.g., architecture patterns), architectural decisions (e.g., technology decisions) from diverse sources, including Q\&A sites and Wiki. Among others, the mined architectural information can help architects evaluate candidate architectural solutions to a given architecture design concern when conducting AE.

\textit{Datasets and sources in AE} - Code-based (e.g., {[S34]}) and text-based (e.g., {[S2]}{[S14]}) datasets are the common datasets in AE activity. On the other hand, Q\&A sites, Wiki, issue tracking systems, and technical blogs and tutorials are the most employed sources in this activity.

\textbf{\textit{Architecture Synthesis (AS)}}: In AS activity, a set of candidate architectural solutions are proposed to address the ASRs that have been gathered during AA, and the AS activity essentially links the problem space to the solution space during architectural design \citep{hofmeister2007general}. The AS activity is the topic of 13.5\% (14 out of 104) of the studies.

\textit{Problems solved and approaches used in AS} - Classification-based approaches were predominately used for mining six types of architectural information to solve the problems related to AS. The problems solved in AS include classification of architectural solutions, architectural decisions, and system requirements. For instance, in {[S4]}, Soliman et al. used a qualitative analysis approach to mine architectural information for technology decisions from Stack Overflow and classified the architectural information into several categories including solution synthesis, in which the mined architectural information in the solution synthesis category provides developers with a set of technology solutions for addressing architectural concerns. 

\textit{Datasets and sources in AS} - Text-based dataset (e.g., {[S15]}{[S79]}) is the frequently used dataset in AS, whereas, Q\&A sites, Wiki, issue tracking systems, and technical blogs \& tutorials are the main sources in this activity.

\textbf{\textit{Architecture Impact Analysis (AIA)}} aims to identify directly affected and indirectly influenced elements of an architecture due to an architectural change scenario \citep{bengtsson2004architecture}. The outcome of this activity aids architects in comprehending the dependencies between the changed elements and the affected elements in the architecture \citep{li2013application}. 13 out of 104 (12.5\%) studies researched on AIA activity.  

\textit{Problems solved and approaches used in AIA} - Similar to the above activities, most of the studies in AIA used classification-based approaches to solve problems related to AIA. Some of the problems solved in this activity include architectural change classification (e.g., [S35] [S40] [S41]) and system requirements classification (e.g., {[S55]}). For instance, Anish et al. {[S55]} developed an ML-based approach for automatically identifying and mining ASRs from requirements documents, and classifying the mined ASRs into several subcategories based on the types of architectural impacts the mined ASRs can have on the system components, which helps architects perform AIA.

\textit{Datasets and sources in AIA} - Text-based (e.g., {[S32]}{[S55]}) and code-based (e.g., {[S35]}{[S88]}) datasets are the common dataset in AIA. Regarding sources, version control systems, Wiki, issue tracking systems, and technical blogs \& tutorials are the prevalent sources in this activity.

\textbf{\textit{Architecture Conformance Checking (ACC)}} is conducted to check and evaluate if the implemented architecture conforms to the intended architecture, and this activity can help architects identify and correct architectural violations and further avoid constant architecture erosion during the software development life cycle \citep{rocha2017preventing}. 9.6\% (10 out of 104) of the studies investigated ACC activity.  
 
\textit{Problems solved and approaches used in ACC} - Few primary studies worked on ACC to solve several problems, including classification/clustering architectural technical debt (e.g., architectural conformance issue), improving architectural description via mining architectural views and models to facilitate ACC. Clustering-based approaches are predominately utilized to solve the mentioned problems in ACC. For example, {[S81]} proposed an architectural conformance-checking approach called Arch-KDM that automatically compares the planned architecture with the current/implemented architecture in order to identify and cluster architectural violations/drifts. 

\textit{Datasets and sources in ACC} - Code-based (e.g., {[S85]}{[S81]}) and text-based (e.g., {[S51]}) datasets are the main datasets used in ACC. Version control systems, issue tracking systems, search engines, and Wiki are the common sources for collecting data in ACC (see the Architecting Activities sheet in the Supplementary Material \citep{dataset}). 

\begin{tcolorbox}[colback=gray!5!white,colframe=gray!75!black,title=Key Findings of RQ3]
\textbf{Finding 3}:
Considerable effort has been devoted to mining architectural information from software repositories to support various architecting activities (i.e., 11 architecting activities). As a result, the current research on mining architectural information can help practitioners be aware of which architecting activities (such as AU and AME) can be supported by the mined architectural information.

\textbf{Finding 4}: 
Although various types of problems (e.g., recovering architectural description via mining architectural views and models) have been addressed in different architecting activities (e.g., ARec), there are still unresolved problems in certain architecting activities, including mining and recommending the implementation of certain architectural elements, such as architectural pattern implementations to assist AIm activity.

\end{tcolorbox}
 
\subsection{Results of RQ4: Approaches and Tools}\label{ResultsOfRQ4}
We describe the approaches and tools used for mining architectural information in Section \ref{Approach} and Section \ref{Tools}, respectively. 

\subsubsection{Approaches}\label{Approach}
The selected studies have proposed and used different approaches to mine architectural information from different sources. We analyzed and synthesized the approaches used for mining architectural information by following the coding steps for data synthesis as described in Section \ref{DataSynthesis}. We observed that some selected studies did name their approaches used to mine architectural information and others did not. If a study does not provide a name for the proposed approach, we labeled it as “Unnamed” (see the Automatic Approaches sheet in the Supplementary Material \citep{dataset}). Moreover, we found that the proposed approaches have different levels of automation (e.g., automatic, semi-automatic). In addition, these approaches have been applied in different types of tasks (e.g., classification, clustering) for mining architectural information from various sources. Thus, in this SMS, we utilized a two-level categorization to categorize the approaches for mining architectural information. At the first level, we divided the approaches in three high-level categories based on the level of automation: (1) \textit{automatic}, (2) \textit{semi-automatic}, and (3) \textit{manual}. At the second level, we further categorized the approaches in three categories based on different types of tasks in mining architectural information: (1) \textit{architectural information classification}, (2) \textit{architectural information clustering}, and (3) \textit{architectural information retrieval}. In total, we gathered 95 approaches from the selected studies of which 60\% (i.e., 57 out of 95) are automatic approaches (see the Automatic Approaches sheet in the Supplementary Material \citep{dataset}), 34.7\% (i.e., 33 out of 95) are semi-automatic approaches (see the Semi-automatic Approaches sheet in the Supplementary Material \citep{dataset}), and 5.3\% (5 out of 95) are manual approaches (see the Manual Approaches sheet in the Supplementary Material \citep{dataset}). Architectural information classification-based approaches count 52.6\% (50 out of 95), followed by clustering-based approaches (26.3\%, 25 out of 95), and information retrieval-based approaches (15.8\%, 15 out of 95). Furthermore, to help and ease the applicability of these approaches in practice, we also present the sources that the proposed approaches can be applied on.  

\textbf{Architectural information classification} refers to the approaches that mine architectural information based on classification techniques. Architectural information classification-based approaches are the most frequently 56.2\% (50 out of 95) used approaches in mining architectural information. Classification techniques learn from the data input provided to them and then use this learning to classify new observations \citep{kotsiantis2006machine}. There are two types of architectural information classification-based approaches that were employed in the selected studies: binary classification-based approaches (e.g., {[S6]}{[S48]}{[S59]}) and multi-class classification-based approaches (e.g., {[S40]}{[S76]}{[S79]}). Architectural information classification-based approaches are applied either automatically, semi-automatically, or manually (see the Automatic Approaches, Semi-automatic Approaches, and Manual Approaches sheets in the Supplementary Material \citep{dataset}). For example, in {[S40]}, Mondal et al. proposed an approach named Archinet for mining and classifying architectural changes from code and commit messages in GitHub. Specifically, they formulated this problem as a multi-class classification task that automatically classifies architectural changes into four categories. In {[S6]}, Bi et al. developed a semi-automatic dictionary-based mining approach to extract QAs and Architectural Tactic (AT) related discussions in Stack Overflow (SO) posts. They formulated this problem as a binary classification task, which automatically identifies QA-AT related discussions from SO posts. Then the authors went on to manually structure the design relationships between QAs and ATs used in practice, and build a knowledge base of how developers use ATs with respect to QA concerns. The proposed approach achieved good performance and F-measure. In {[S6]}, Malavolta et al. employed a manual approach (i.e., qualitative analysis) to mine and elicit evidence-based architectural guidelines for open-source ROS-based software from GitHub, BitBucket, and GitLab. Their qualitative data analysis yielded 39 guidelines for architecting robotics software, which can be used by roboticists to architect their systems to achieve particular quality requirements.

\textbf{Architectural information clustering} denotes the approaches that employ clustering-based techniques to mine architectural information from various sources. Architectural information clustering-based approaches count 26.3\% (25 out of 95 approaches). The clustering techniques aim at reducing the amount of data by categorizing or grouping similar data items together into subsets or clusters \citep{saxena2017review}. The goal is to create clusters with internal coherence, placing similar objects in the same group, and assigning dissimilar objects to different groups. In this sense, documents that belong to a certain cluster should be as similar as possible and dissimilar from documents in other clusters. We identified two types of architectural information clustering-based approaches employed in the selected studies: partitional clustering (e.g., {[S43]}) and hierarchical clustering (e.g., {[S68]}). In addition, these approaches operate in two ways: automatic and semi-automatic (see the Automatic Approaches and Semi-automatic Approaches sheets in the Supplementary Material \citep{dataset}).

Partitional clustering approaches output an initial partition with a certain number of clusters. Specifically, partitional clustering approaches divide a dataset into a number of groups based upon a certain criterion known as fitness measure \citep{nanda2014survey}. The fitness measure directly affects the nature of the formation of clusters. Once an appropriate fitness measure is selected, the partitioning task is converted into an optimization problem (e.g., grouping based on minimization of distance or maximization of correlation between patterns) \citep{nanda2014survey}. For example, Casamayor et al. {[S43]} used a partitional clustering-based approach to semi-automatically mine the potential responsibilities of components in a software system to be developed. Specifically, NLP techniques and the K-means algorithm have been applied to automatically cluster candidate responsibilities into groups. Firstly, this approach processes requirements documents by the Part-Of-Speech (POS) tagging technique to detect actions, activities, or tasks that will become responsibilities of certain components in the architecture of a system. Afterward, K-means is utilized to group (i.e., partitional clustering) similar responsibilities into architectural components. The proposed approach achieved good performance in terms of precision and recall. 

Hierarchical clustering approaches work by iteratively merging smaller clusters into larger ones, or by splitting larger clusters onto small ones \citep{saxena2017review}. The key point is the rule used by the approach to decide which small clusters are to merge or which larger clusters are to split. The result is a tree of clusters showing how clusters are related. The output of hierarchical clustering is a hierarchy, a structure that is more informative than an unstructured set of clusters \citep{saxena2017review}. For instance, in {[S68]}, Mitchell and Mancoridis presented an approach that automatically searches and hierarchically groups highly interdependent modules into the same subsystems/clusters and, conversely, groups independent modules into separate subsystems/clusters. The approach was developed to help software engineers perform a variety of large and complex system understanding and maintenance activities.

\textbf{Architectural information retrieval} covers the approaches that employ Information Retrieval (IR) based techniques \citep{singhal2001modern} to search and retrieve architectural information in various sources. Around (15.8\%, 15 out of 95) of the approaches are based on IR techniques, which have been utilized to (semi-)automatically search and retrieve architectural information by using two types of searching mechanisms, namely keyword-based search and semantic search. 

On the one hand, keyword-based search approaches do not understand polysemy and synonymy \citep{guha2003semantic}. Specifically, when looking at a repository, a keyword-based approach looks for the distribution of documents within the repository to find how relevant a document is to the search query of the user \citep{guha2003semantic}. Basically, this means that a document with similar words/terms to those the user types into a search engine will be thought to be more relevant and will appear at a higher position in the search results \citep{guha2003semantic}. For example, Aman-ul-haq and Ali Babar {[S23]} proposed a tag-based/annotation-based knowledge identification and extraction approach. One of the functionalities of this approach is the keyword-based search functionality, which can search and retrieve desired architectural artifacts in the repository using the keywords attached to each architectural artifact. This approach has been developed to reduce the time and effort required for searching and capturing architecture knowledge in a software repository. 

On the other hand, semantic search-based approaches understand polysemy and synonymy and know the meaning of the words \citep{guha2003semantic}. Specifically, semantic search-based approaches are designed to understand the context the words are used within the documents in order to match these words/terms more accurately to the user search queries \citep{guha2003semantic}. For instance, in {[S13]}, Mujhid \textit{et al.} presented an approach based on IR and program analysis techniques for searching and reusing architectural tactics implemented in OSS projects. Among other algorithms used by this approach, the approach uses a novel information retrieval algorithm that automatically searches and ranks the source files implementing architectural tactics in the software repositories based on (i) the semantic similarity of a source file to a searched architectural tactic and (ii) the semantic similarity of a source file and its direct dependent files to a technical problem represented in the search query. The approach was developed to help developers search and reuse implementation examples of architectural tactics for a given technical context. Technical context refers to a framework, programming language, or API that can be used to implement the tactic {[S13]}. 

%
%

\subsubsection{Tools}\label{Tools}
Various tools are proposed to facilitate searching, extracting, and mining architectural information, and make it more convenient for developers to fasten the development process. We explored the selected studies and collected the tools that support mining architectural information. In total, 56 tools were collected from the selected studies. We categorized these tools into two categories: \textit{general tools} 30.4\% (17 out of 56) and \textit{dedicated tools} 69.6\% (39 out of 56).

General tools are provided to analyze, process, and mine various information, not limited to architectural information. For example, Stanford Parser is a general NLP tool frequently used in textual information prepossessing tasks, and Gokyer \textit{et al}. {[S66]} used Stanford Parser for analyzing the grammar of textual architecture information in Wiki. We provide the collected general tools, their descriptions, URL links, and the relevant studies in which these general tools were used in the General Tools sheet in the Supplementary Material \citep{dataset}.

Dedicated tools are developed for specifically mining architectural information. We found that dedicated tools have two levels of automation (i.e., automatically and semi-automatically) and have been applied in different types of tasks (e.g., clustering) for mining architectural information from various sources. Thus, similar to what we did when analyzing architectural information mining approaches, we utilized a two-level categorization to categorize the dedicated tools for mining architectural information. At the first level, we divided these tools in three high-level categories based on the level of automation: (1) \textit{automatic} (82\%, 32 out of 39 tools) and (2) \textit{semi-automatic} (17.9\%, 7 out of 39 tools) (see the Automatic Tools and Semi-automatic Tools sheets in the Supplementary Material \citep{dataset}). At the second level, we further categorized the dedicated tools in three categories based on different types of tasks in mining architectural information: (1) \textit{architectural information classification} (35.9\%, 14 out of 39 tools), (2) \textit{architectural information clustering} (38.5\%, 15 out of 39 tools), and (3) \textit{architectural information retrieval} (25.6\%, 10 out of 39 tools). Moreover, to help and ease the applicability of these approaches in practice, we also present the sources that the proposed approaches can be applied on. The detail information about the dedicated tools is shown in the Automatic Tools and Semi-automatic Tools sheets in the Supplementary Material \citep{dataset}. Specifically, these sheets list the categories of different tasks in mining architectural information, tool names, sources used, mined architectural information, URL links, and the relevant studies in which these dedicated tools were proposed. Note that, the results in the “URL link” column were collected from the selected studies. Moreover, it is worth noting that some URL links of the tools were not accessible when we conducted this SMS. Therefore we provide the links of these tools with a note “Not accessible”. On the other hand “Not provided” means that the selected study does not provide a URL link to the proposed tool. 


As mentioned above, many approaches and tools were proposed for mining architectural information. However, there is no clear answer as to which architecting activities the proposed approaches and tools can be best applied and how to leverage the suitable approaches and tools for these activities. Answering these questions would be beneficial for both practitioners and researchers to make informed decisions to solve a problem or conduct research using architectural information mining approaches and tools. Since validation holds the key to see the applicability of an architectural information mining approach or tool, we examined if and how the proposed approaches and tools were validated. To achieve this, we checked the evidence reported in the primary studies to synthesize the evidence levels on the approaches and tools. Specifically, we investigated the evidence level of each study related to the applicability of the approaches and tools in practice. To do so, we adopted the evidence hierarchy proposed in~\citep{alves2010requirements}. The evidence levels are defined from weakest to strongest. For example, Evidence Level 6 indicates that the proposed approach or tool has already been adopted by industrial organizations for daily engineering practice.

\begin{itemize}
    \item Level 1: No evidence.
    \item Level 2: Evidence obtained from demonstration or working out with toy examples.
    \item Level 3: Evidence obtained from expert opinions or observations (e.g., surveys or interviews from practitioners).
    \item Level 4: Evidence obtained from academic studies (such as controlled lab experiments conducted in an academic setting, e.g., with students).
    \item Level 5: Evidence obtained from industrial studies (such as causal case studies conducted in an industrial setting, e.g., with practitioners).
    \item Level 6: Evidence obtained from industrial practice (e.g., the approach or tool has already been adopted by industrial organizations).
\end{itemize}

\begin{figure}
   \centering
   \subfloat[Distribution of the proposed approaches across evidence levels]{\includegraphics[width=.5\linewidth]{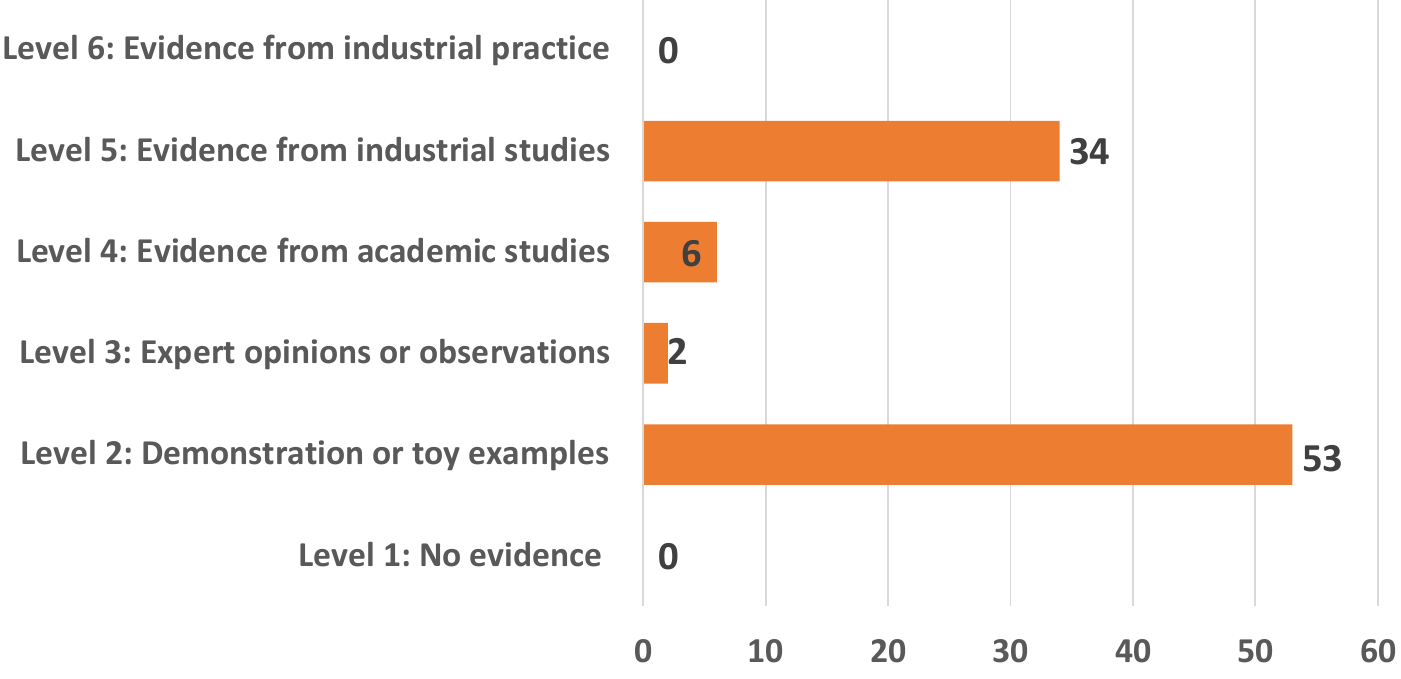}}\hfill 
    \subfloat[Distribution of the proposed tools across evidence levels]{\includegraphics[width=.5\linewidth]{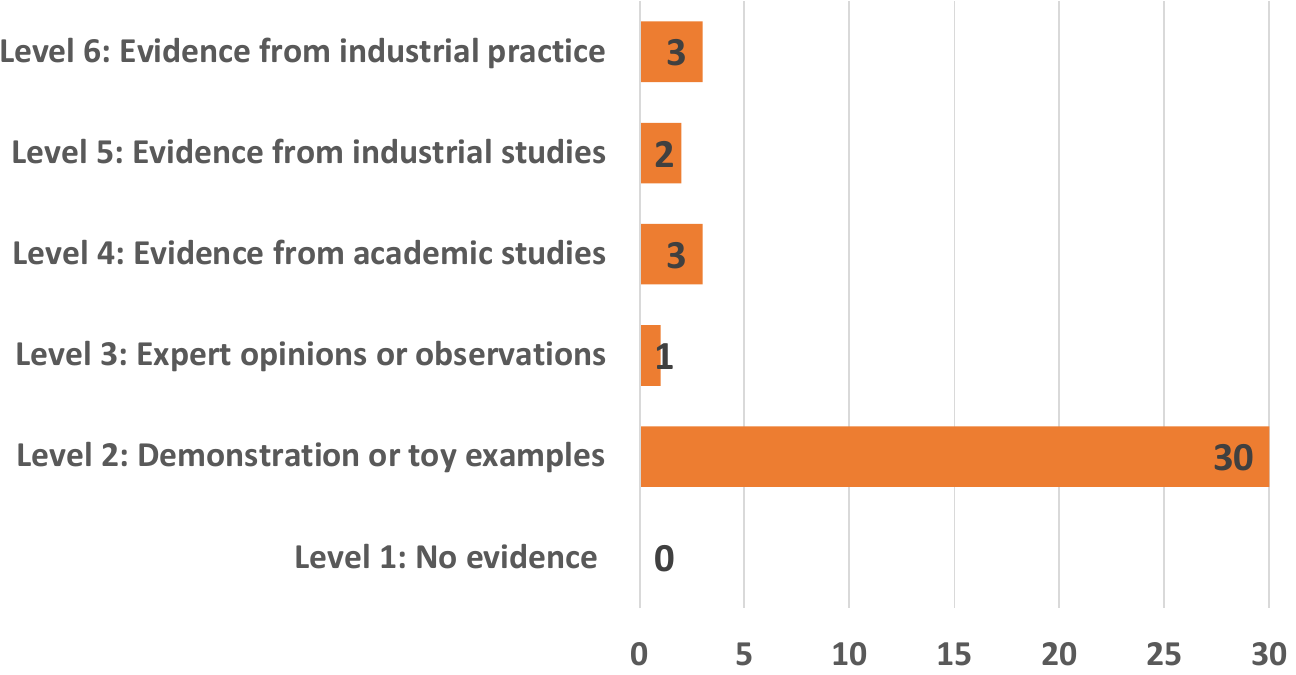}}\hfill
    \caption{Distribution of the proposed approaches and tools across evidence levels}
    \label{EvidenceLevels}
\end{figure}

The high evidence level can provide practitioners sufficient confidence to employ architectural information mining approaches and tools in architecting activities. Figure 10(a) and Figure 10(b) show the distribution of the proposed approaches and tools across the six levels of evidence, respectively. We can see that demonstration and toy examples (Level 2, 53 out of 95, 55.8\%) is the most common evidence used to evaluate architectural information mining approaches. In most of these studies (e.g., [S1][S6][S39][S60][S71]), the authors first introduced the details of the proposed mining approaches or tools that support architecting activities and then applied these approaches and tools in some cases to show their applications. For example, in {[S71]}, Diaz et al. proposed a forward-looking approach that is able to infer groups of likely module dependencies that can anticipate architectural smells in a future version of the system.

6 out of 95 (6.3\%) studies (e.g., {[S13]}{[S16]}) have conducted academic studies (e.g., academic controlled experiments) to evaluate their approaches (Level 4). 2 out of 95 (2.1\%) studies (i.e., {[S24]}{[S61]}) reported expert insights and observations to illustrate how the proposed approaches can support architecture design through mining architectural information (Level 3). 34 out of 95 (35.8\%) studies (e.g., [S53][S45][S55][S64]) conducted industrial studies (e.g., industrial case studies) to validate the architectural information mining approaches (Level 5). 
For the proposed approaches, there is no study that provides evidence from industrial practice (Level 6), which shows that the proposed architectural information mining approaches have not been evaluated in industry practice. on the other hand, only three architectural information mining tools (i.e., [S19][S37][S99]) have been adopted in industrial practice (i.e., Level 6). Furthermore, the majority of the tools (e.g., [S23] [S54][S66]) have been evaluated through the demonstration and toy examples (see Figure 10(b)). 
It is worth noting that [S64] used both demonstration examples and industrial case studies to provide concrete evidence about the effectiveness of the proposed approach. In this case, we chose the higher one (i.e., industrial study, Level 5) as the evidence level.




\begin{tcolorbox}[colback=gray!5!white,colframe=gray!75!black,title=Key Findings of RQ4]
\textbf{Finding 5}:
The primary studies employed three types of approaches (automatic, semi-automatic, and manual approaches) to mine architectural information from software repositories. Moreover, since manual approaches are time-consuming and require substantial human efforts, most primary studies motivated their mining approaches to reduce the manual effort in mining architectural information, where 60.0\% (i.e., 57 out of 95) are automatic approaches and 34.7\% (i.e., 33 out of 95) are semi-automatic approaches, and these approaches were used to solve various tasks, such as architectural information classification and clustering.  

\textbf{Finding 6}:
Although various architectural information approaches (i.e., 95 approaches) have been proposed for mining architectural information from software repositories, we found that still more than half of the approaches (i.e., 56 approaches) are not supported by tools.
 
\end{tcolorbox}

\subsection{Results of RQ5: Challenges}\label{ResultsOfRQ5} 
The challenges refer to the obstacles or issues encountered in mining architectural information. Our data synthesis yielded four categories, of which one was categorized as “Others” (referring to codes that do not fit into the already generated subcategories), and seven subcategories. Table \ref{Challenges} shows these categories and subcategories, their descriptions along with relevant studies.

\small
\begin{longtable}{p{6em}p{14em}p{12em}p{0.1em}}
\caption{Categorization of the challenges encountered in mining architectural information} \label{Challenges} \\\hline
\multicolumn{1}{l}{\textbf{Type}} & \multicolumn{1}{l}{\textbf{Subtype}} & \multicolumn{1}{l}{\textbf{Studies}} & \multicolumn{1}{l}{\textbf{Count}}  \\\hline  
\endfirsthead
\multicolumn{4}{c}
{{\bfseries \tablename\ \thetable{} -- continued from previous page}} \\\hline 
\multicolumn{1}{l}{\textbf{Type}} & \multicolumn{1}{l}{\textbf{Subtype}}  & \multicolumn{1}{l}{\textbf{Studies}} & \multicolumn{1}{l}{\textbf{Count}} \\\hline  
\endhead
\hline \multicolumn{4}{r}{{Continued on next page}} \\ 
\endfoot
\hline
\endlastfoot
                                     
{Architectural element description (26)} 
                                                    & Vagueness or ambiguity in architectural element description 
                                                    & {[S1]} {[S15]} {[S11]} {[S18]} {[S20]} {[S40]} {[S42]} {[S45]} {[S63]} {[S72]} {[S77]}  
                                                    &  {11} \\ \cline{2-4} 
                                                     
                                                    & Incompleteness in architectural element description  
                                                    & {[S1]} {[S6]} {[S10]} {[S15]} {[S40]} {[S48]} {[S58]} {[S63]} {[S72]}
                                                    &  {9} \\ \cline{2-4} 
                                                     
                                                    & Redundancy in architectural element description 
                                                    & {[S40]} {[S45]} {[S58]} {[S63]} {[S72]} {[S83]}   
                                                    &  {6} \\ \cline{1-4} 

{Approach and tool (23)} 
                                                   & Limitation of approaches and tools
                                                   & {[S27]} {[S30]} {[S59]} {[S64]} {[S88]} {[S89]} {[S90]} [S91] {[S94]} {[S97]} {[S100]} {[S101]} {[S102]} {[S104]} 
                                                   &  {14} \\ \cline{2-4}
                                                   
                                                   & Applicability of approaches to heterogeneous architectural data
                                                   & {[S17]} {[S24]} {[S25]} {[S27]} {[S32]} {[S48]} {[S64]}  
                                                   &  {7} \\ \cline{2-4}
                                                   
                                                   & Feature selection
                                                   & {[S5]}{[S20]}   
                                                   &  {2} \\ \cline{1-4}

{Architectural dataset (17)}            
                                                    & Size of architectural dataset for ML model training
                                                    & {[S4]} {[S13]} {[S15]} {[S25]} {[S40]} {[S45]} {[S50]} {[S55]} {[S64]} {[S72]} 
                                                    &  {10} \\ \cline{2-4} 
                                                    
                                                    & Quality of architectural dataset for ML model training  
                                                    & {[S1]} {[S6]} {[S13]} {[S14]} {[S15]} {[S24]} {[S40]} 
                                                    &  {7}  \\ \cline{1-4}                                                   
                                            
{Others (10)} 
                                                   & 
                                                   & {[S10]} {[S13]} {[S14]} {[S28]} {[S34]} {[S43]} {[S46]} {[S48]} {[S55]} {[S59]} 
                                                   &  {10} \\ \cline{1-4}
       
\end{longtable}				
\normalsize

\textit{\textbf{Architectural element description}}: As shown in Figure \ref{SourceMinedFigure} (the results of RQ2 in Section \ref{ResultsOfRQ2}), the selected studies used diverse sources for mining architectural information. However, the description of architectural elements in these sources is critical for architectural information mining approaches and tools. 25\% (26 out of 104) of the studies reported the challenges related to architectural element descriptions encountered when mining architectural information. We further categorized these challenges into three subcategories (see Table \ref{Challenges}).

\textit{Vagueness or ambiguity in architectural element description}: Architectural elements, including architectural patterns, architectural components, and architectural rationale (e.g., benefits and drawbacks of architectural solutions) are not clearly described in certain sources, which leads to challenges on architectural information mining approaches (e.g., manual or automatic approaches) and tools. 10.6\% (11 out of 104) of the studies reported the challenge related to the vagueness or ambiguity in architectural element descriptions in different sources of software repositories. For example, in {[S1]}, Soliman \textit{et al}. proposed a semi-automatic approach for mining AK (including architectural components, architectural rationale, and system requirements) from issue tracking systems (e.g., Jira). However, the authors argued that it is quite challenging to accurately identify and mine AK concepts in issue tracking systems due to, for example, vagueness and implicitness in architectural concept descriptions in those sources. For example, the authors stated that architectural concepts which are described vaguely or implicitly rend them to be difficulty identified and mined inaccurately, since the meanings of the architectural concepts are unclear and it may not be determined from their contexts, and there also might be more than one meanings in the description of those architectural concepts.

\textit{Incompleteness in architectural element description}: Some architectural elements are incompletely described in certain sources, and this may cause various problems (e.g., less accuracy of ML-based approaches) in mining architectural information. 8.7\% (9 out of 104) of the studies reported the problem related to the incompleteness in architectural element descriptions in various source of software repositories. For example, Karthik and Medvidovic {[S20]} proposed a DL-based approach to automatically mine inter-component relationships from Stack Overflow posts. However, the authors reported that due to the incompleteness in architectural component descriptions in Stack Overflow posts, the approach can include irrelevant architectural components and exclude relevant information (such as version information of architectural components). Furthermore, Ven \textit{et al}. {[S18]} claimed that it is difficult to identify and mine the made architectural decisions in commit messages because sometimes architecture decisions in commit messages are cryptic, short, or incomplete. 

\textit{Redundancy in architectural element description}: Redundant descriptions in the architectural datasets induce redundant features. Empirical evidence from features selection literature shows that, along with irrelevant features, redundant features also affect the performance of Deep Learning (DL)/ML algorithms and thus should be eliminated as well \citep{yu2003feature}. Moreover, redundant descriptions lead to performing the same operations repeatedly, which is labor-intensive and time-consuming in mining architectural information. 5.8\% (6 out of 104) of the studies mentioned the challenge related to the redundancy in architectural element descriptions in various source of software repositories. For example, in {[S45]}, Lian \textit{et al}. proposed an ML-based approach that is based on NLP and clustering techniques for assisting architects to identify requirements knowledge on components from a collection of domain documents. However, the authors reported the challenges related to redundant descriptions of various architectural elements (e.g., system components and features) in those documents.  

\textbf{\textit{Approach and tool}} denotes the challenges related to approaches or tools used in mining architectural information, and these challenges were reported in 22.1\% (23 out of 104) of the studies. 

\textit{Limitation of approaches and tools}: The results highlight that 13.5\% (14 out of 104) of the studies considered the limitation of architectural information mining approaches and tools as a serious challenge. For example, in {[S27]}, Gopalakrishnan \textit{et al}. proposed an approach that could only mine architectural tactics from project source code. However, this approach is limited to mining nine architectural tactics (such as, audit trail, authentication, heartbeat). To mine additional architectural tactics, the approach will need to be expended, which will require performing additional training based on source code that implements other tactics.

\textit{Applicability of approaches to heterogeneous architectural data}: As pointed out by \cite{canfora2015defect}, most software projects are heterogeneous (e.g., in terms of size, domain specifications, architecture patterns, programming languages, and software models). Therefore, due to this heterogeneity, specific architectural information mining approaches and tools can hardly be generalized or applied to mine architectural information from various architectural data. 6.7\% (7 out of 104) of the studies reported the problem related to the applicability of the proposed approaches to different architectural data. For example, in {[S48]}, Mahadi \textit{et al}. proposed an ML classification based approach for mining design discussions in several sources (including VCS (e.g., GitHub), issue tracking systems (e.g., Jira), and Q\&A sites (Stack Overflow)). However, the authors reported the challenges related to poor performance (i.e., accuracy) when applying this ML classifier trained on the architectural dataset from one source (e.g., GitHub) to a different architectural dataset from another source (e.g., Stack Overflow).

\textit{Feature selection}: As many candidate features are introduced to better represent various research domains, diverse studies extracted and used a greater variety of features to train architectural information mining approaches. However, not all features used are beneficial to improving the performance (e.g., accuracy) of architectural information mining approaches. First, some interdependent features do not fit when applied to architectural information mining approaches. Another reason is that using too many features may result in overfitting issues \citep{yang2022predictive}. Therefore, how to select the most suitable features for mining architectural information has become a critical challenge. 
In {[S20]}, Karthik and Medvidovic proposed an approach that uses noun chunking to identify inter-component relationships from Stack Exchange website. However, this approach can exclude some relevant information from the noun chunk, and sometimes words unrelated to architectural components get included as of the noun chunk. The authors stated that a supervised ML approach for the Named Entity Recognition problem could be used to improve the accuracy.

\textbf{\textit{Architectural dataset}} is a critical asset for several architectural information mining approaches and tools to extract real and significant architectural information from the datasets. However, there are no public and suitable datasets for certain architecture design tasks. Architectural dataset is reported as a challenge in 16.3\% (17 out of 104) of the studies. We further categorized the challenges in the architectural dataset category into two subcategories (see Table \ref{Challenges}). 

\textit{Size of the architectural dataset for ML model training}: The size of dataset is a matter, specifically, the size of dataset for training ML-based approaches in mining architectural information, since it can influence the performance (e.g., accuracy rate) of these approaches. 9.6\% (10 out of 104) of the studies that used ML-based approaches reported the challenge related to the size of architectural dataset in mining architectural information. For example, Gilson \textit{et al}. {[S42]} proposed an ML-based approach for extracting and mining QAs from user stories for early architecture decisions making. However, the authors mentioned that the approach did not perform well due to the size of dataset (i.e., the dataset is too small) used in the training of ML models, and thus the approach needs some improvements with large datasets.  

\textit{Quality of the architectural dataset for ML model training}: The architectural datasets vary much in quality, including bias, noise, imbalance, and mislabelling. There are many factors that may impact the quality of architectural datasets, such as the source of the dataset, whether the dataset has been preprocessed, and the ground truth in the dataset \citep{yang2022survey}, and any of these factors will influence the performance of the approaches or tools used. Quality of the architectural dataset is the challenge of 6.7\% (7 out of 104) of the selected studies. In {[S24]}, L\'{o}pez \textit{et al}. proposed an approach named Toeska Rationale Extraction (TREx) for extracting, recovering, and exploring architectural rationale information from various sources, such as Wiki, meeting notes, emails. However, the authors stated that the effectiveness (e.g., accuracy) of the approach on extracting data (e.g., architectural rationale) is heavily dependent on the quality of the available textual architectural artifacts.

\textbf{Others} refer to other factors (such as time and labor constraints) that could influence the effectiveness of architectural information mining approaches. For example, in {[S43]}, Casamayor \textit{et al}. presented a text mining approach for analyzing textual requirements in order to obtain a functional decomposition of a system in responsibilities, that can guide the design of architecture. However, the authors reported that it required a lot of time and effort to manually annotate textual data in order to accomplish this approach.
 
\begin{tcolorbox}[colback=gray!5!white,colframe=gray!75!black,title=Key Findings of RQ5]
\textbf{Finding 7}:
The challenges in mining architectural information are mainly derived from the descriptions of architectural elements, wherein vagueness or ambiguity in architectural element description is a prevalent challenge. 

\textbf{Finding 8}: The primary studies also report the challenges related to the limitation of architectural information mining approaches and tools. For example, there is still a lack of effective architectural information mining approaches and tools that are applicable to various architectural datasets from diverse sources in mining architectural information. Moreover, the number of quality datasets with an appropriate size available for architectural information mining approaches is also a serious challenge.

\end{tcolorbox}

             
             

\section{Discussion}\label{Discussion}

\subsection{Analysis of Results}\label{AnalysisResults}

\subsubsection{Mined Architectural Information Could Support Architecting Activities}\label{Discussion_RQ1 and RQ3}

The results of RQ1 (see Section \ref{ResultsOfRQ1}) reveal that various categories and subcategories of architectural information have been mined in software development, such as \textit{architectural description}, \textit{architectural solution}, and \textit{system requirement} (see Table \ref{minedArchitecturalInfo}). Moreover, the results show that \textit{architectural description} is the most mined category of architectural information. The potential reason could be that architectural description plays as a blueprint for system development, specifically a blueprint for effectively and successfully developing large and complex systems \citep{clements2003documenting}. In addition, the information in architectural description, such as architectural views, which describe systems in multiple views (e.g., process view and development view), enables the architecture to be communicated to, and understood by, different stakeholders. Also, it allows stakeholders to verify whether the architecture will address their concerns \citep{6129467}. The core benefits of mining architectural description are lowering architectural knowledge vaporization and speeding up maintenance of software systems \citep{borrego2019towards}. These benefits can be achieved by capturing, sharing, and reusing architectural information (such as architectural views, architectural models, and architectural rationale) in the architectural description, thus reducing the time that new developers need to familiarize themselves with an existing system and potentially speeding up the development. \textit{Architectural solution} is the second most mined category of architectural information. The reason could be that researchers and practitioners consider architectural solutions, such as architectural patterns (e.g., service-oriented architecture) and tactics (e.g., resource pooling) as the fundamental building blocks in the architecture design of a system, and they are used to address ASRs \citep{SA2012}. Contrarily to changing implementation (e.g., low-level source code), once an architecture solution (e.g., an architecture pattern) is adopted and implemented, it is quite difficult and costly to change it \citep{SA2012}. Mining architectural solutions promises significant practical benefits including increased reuse of architectures, lower costs of architectural change, and less erosion of architectural design. These benefits are obtained by offering developers access to the architectural solutions that lead to the architecture, rather than only the resulting outcomes and artifacts from the architectural solutions \citep{kruchten2004ontology}.

The results of RQ3 show that eleven architecting activities can be supported by the mined architectural information, where \textit{architecture understanding} is the most supported activity (i.e., 44.2\% (46 out of 104)). Some mined architectural information to support AU include architectural description, architectural decision, system requirement, and architectural solution (see the Architecting Activities sheet in the Supplementary Material \citep{dataset}). An architecture of a system should be well understood before any corrections or changes can be applied to it. AU is a prerequisite for and supportive to other architecting activities (e.g., AA, AI, AME). Architectural elements, such as architectural decisions, are vital in achieving the desired software quality, and a good understanding of these architectural elements and their relationships is important in supporting AU \citep{stevanetic2014exploring}. Understanding architectural decisions that make an architecture is also a key factor for maintaining the software system \citep{shahin2014architectural}. For example, AU enables maintainers to gain a comprehensive overview of the complex dependency relationships between architectural components and connectors. Thus, these facts tend to motivate researchers and mine several types of architectural information from software repositories in order to facilitate AU during the architecting process. Architecture maintenance and evolution is the second most supported activity. It is an established fact that good software architecture and design leads to better evolvability, maintainability, availability, software cost reduction, and so on \citep{SA2012}. However, during development, developers tend to willingly or unwillingly deviate from the architectural patterns through compromising architecture and design process by poor or hasted architectural design decisions, which lead the architecture to different architectural problems or anomalies (e.g., architectural smells). As the system evolves, the accumulation of such problems can cause the implemented architecture to deviate away from the intended architecture, i.e., a typical case of architecture erosion~\citep{li2022understanding}. An architecture with such problems can aggravate the brittleness of the system \citep{perry1992foundations}, decrease architecture sustainability \citep{koziolek2013measuring}, and make it difficult for developers to understand the internal structure of the system \citep{perry1992foundations}. Therefore, these facts motivate researchers to turn their attention to developing architectural information mining approaches and tools to assist AME activity.

Furthermore, sometimes architectural information is tacit and not documented at all \citep{ding2014open}. With the increase in the size and complexity of the systems, architectural information management becomes more challenging, and stakeholders need to find the right information and recover it efficiently. Searching and recovering relevant architectural information from various artifacts is a challenging task. With a large volume of artifacts, applying mining approaches can greatly facilitate the task of recovering desired architectural information. Thus, it is not surprising that \textit{architecture recovery} is the third (i.e., 32.7\% (34 out of 104)) most supported architecting activity (see the Architecting Activities sheet in the Supplementary Material \citep{dataset}). Architecture can be discovered from various artifacts, such as source code and software documentation \citep{ducasse2009software}. During our data synthesis, we found that most of the studies supporting AR recovered architectural information, such as architectural description (e.g., architectural views and models) from source code (e.g., {[S39]}{[S47]}). 

\textbf{The relationships between the mined architectural information (results of RQ1), sources (RQ2), and supported architecting activities (results of RQ3)}: We show the relationships between mined architectural information, sources, and supported architecting activities through the results of RQ1 (see Section \ref{ResultsOfRQ1}), RQ2 (see Section \ref{ResultsOfRQ2}), and RQ3 (see Section \ref{ResultsOfRQ3}) (see the Architecting Activities sheet in the Supplementary Material \citep{dataset}). This information can, for example, help researchers and practitioners know which architecting activity (e.g., architecture maintenance and evolution) can be supported by what types of mined architectural information (e.g., architectural technical debt) during the development. In addition, such information can also help practitioners know which sources store certain architectural information or elements (e.g., reusable architectural elements) for certain architecting activities (e.g., architecture reuse).

\subsubsection{Various Approaches and Tools are Used to Mine Architectural Information from Diverse Sources}\label{Discussion_RQ4 and RQ2}

The results of RQ4 show that various approaches and tools (see Section \ref{Approach} and Section \ref{Tools}) have been proposed and employed to mine architectural information from diverse sources (results of RQ2, see Figure \ref{SourceMinedFigure}), for example, \textit{VCS} (e.g., GitHub), \textit{Wiki}, \textit{Issue tracking systems}, and \textit{Chat messages}. As shown in Figure \ref{SourceMinedFigure}, \textit{VCS} (54.8\%, 57 out of 104 studies) is the most frequently used source for mining architectural information followed by \textit{Online Q\&A sites} (31.7\%, 33 out of 104 studies). One reason is that these two sources are pervasively used in software development. For example, tens of millions of software practitioners use VCS, such as GitHub, by committing source code and updating related artifacts to the \textit{VCS}, and \textit{Online Q\&A sites} (e.g., Stack Overflow) is the most visited online developer community to search and share software development related information, including architectural information \citep{de2022developers}. Thus, these facts make researchers and practitioners tend to first rely on these two sources and find out valuable data on \textit{VCS} and \textit{Online Q\&A sites} when mining architectural information. Moreover, the proposed approaches and tools have been experimented on different types of datasets for mining architectural information (see Figure \ref{DatasetMinedFigure}) to support various architecting activities, where code-based and text-based datasets are the most common types of datasets. One possible reason is that code-based and text-based datasets are usually the major datasets available in open-source repositories. Few studies conducted experiments on image-based and metric-based datasets due to the lack of such datasets in software repositories.

Regarding the proposed and employed approaches, we found that the research on mining architectural information mainly focuses on automatically mining architectural information from various sources (i.e., 60\%, 57 out of 95 approaches). On the other hand, the results show that 56 tools have been proposed and utilized for mining architectural information, of which 30.4\% (17 out of 56) are general tools and 69.6\% (39 out of 56) are dedicated tools. The general tools are usually employed to support the dedicated tools in mining architectural information. Architectural information classification-based approaches 52.6\% (50 out of 95) and dedicated tools that support architectural information classification (35.9\%, 14 out of 39 tools) are the most frequently used approaches and tools in mining architectural information. One potential reason could be that, classification based approaches and tools normally require an existing classification, which is the common situation in the scenarios of mining architectural information, like the ISO 42010:2011 standard \citep{6129467} for the classification of architecture elements and the architectural decision ontology~\citep{kruchten2004ontology} for the classification of architectural decisions. Moreover, we observed that many dedicated tools (see the Automatic Tools and Semi-automatic tools sheets in the Supplementary Material \citep{dataset}) used to mine architectural information are implemented to support architectural information mining approaches (see the Automatic Approaches, Semi-automatic Approaches, and Manual Approach sheets in the Supplementary Material \citep{dataset}). However, for the proposed architectural information mining approaches (i.e., 95 approaches), still more than half of the approaches (i.e., 56 approaches) are not supported by tools.

\textbf{The relationships between the approaches and tools (results of RQ4) and sources used (results of RQ2) for mining architectural information}: We show the relationships between the tasks (e.g., classification tasks) in mining architectural information, architectural information mining approaches and tools, sources used for mining architectural information, and mined architectural information in the Automatic Approaches, Semi-automatic Approaches, Manual Approach, Automatic Tools, and Semi-automatic Tools sheets in the Supplementary Material \citep{dataset}). These sheets can provide suggestions to researchers and practitioners about which categories of architectural information (e.g., architectural change, design relationship) can be mined by which architectural information mining approaches and tools from which sources (e.g., App Stores, developer mailing lists) of software repositories.  

\subsection{Implications for Researchers and Practitioners}\label{ImplicationsResearchersAndPractitioners}

\subsubsection{Addressing unresolved problems in architecting activities}

Although various types of architectural information have been mined (results of RQ1) to solve the problems in different architecting activities (results of RQ3), there are still unresolved problems in certain architecting activities, including architectural implementation (AIm) and maintenance \& evolution (AME). For example, we found that a few studies focused on mining and recommending the implementation of certain architectural elements, such as architectural tactic implementations (e.g., {[S13]}{[S49]}) in order to assist AIm activity. 
Furthermore, we also noticed that there is a lack of studies on the implementation of other architectural elements, including architectural pattern and framework implementations. 

\textbf{Suggestion 1}: \faHandORight \hspace{0.5mm} We suggest that future research could propose approaches and tools for resolving the unresolved problems in certain architecting activities. For example, we observed that most of the studies (e.g., {[S13]}{[S49]}) in AIm activity employed approaches that rely on traditional information retrieval and ML or program analysis-based techniques for mining and recommending code snippets that implement architectural elements (e.g., tactics [S13]). However, as reported by the authors of [S13], these approaches suffer from the common accuracy problems of automated text mining approaches. Given the considerable success of Large Language Models (LLMs) in software engineering and the abundance of publicly available datasets, such as source code in software repositories, neural information retrieval techniques and LLMs are becoming increasingly popular and offer an attractive alternative to traditional techniques when searching and recommending development artifacts, such as source code \citep{dinh2024large}. Thus, neural information retrieval techniques and LLMs could be potential solutions for resolving unresolved problems in certain architecting activities. If the goal is to mine and recommend architectural relevant code that implements an architectural pattern, one possible direction is to employ neural information retrieval techniques \citep{guo2020deep} and LLMs \citep{chang2023survey} to search and detect an architectural pattern and identify the technical context in which the pattern has been used. To clarify, such an approach could have several components. For example, the first component can be a source code repository, which relies on a code crawler to extract open-source projects from various software repositories, such as GitHub, Krugle, and Koders. The second component could rely on neural information retrieval techniques to represent source code and their source files in the form of an index. This component can support the search and retrieval of code snippets that implement an architectural pattern from the code repository. The third component could be an architectural pattern detector, which can rely on LLMs to detect and clarify various architectural patterns in the indexed code artifacts. The fourth component can be a dependency analyzer, which can rely on static analysis techniques or tools (such as Arcan\footnote{\url{https://essere.disco.unimib.it/wiki/arcan/}}, Dependency Finder\footnote{\url{https://depfind.sourceforge.io/}}) for generating a dependency matrix for each architectural component that constitutes the architectural pattern in the source code of a project. Lastly, the fifth component can use the dependency matrix to find whether the implementation of a given architectural pattern is related to a technical problem/context. The technical context refers to a framework, technology, programming language, or API that can be used to implement the architectural pattern. In other words, it is the technical or design context in which the architectural pattern needs to be implemented. 

\subsubsection{Promising approaches and tools for mining architectural information}
Despite numerous approaches and tools proposed for mining architectural information (see Section \ref{ResultsOfRQ4}), there is no clear answer as to which architecting activities the proposed mining approaches and tools can be best applied and how to leverage the suitable approaches and tools for these activities. Answering these questions would be beneficial for both practitioners and researchers to make informed decisions to solve problems or conduct research using architectural information mining approaches and tools. Thus, based on the analysis conducted in this study, in Table \ref{SuggestingApproachesToPractitioners} and Table \ref{SuggestingToolsToPractitioners}, we provide the most promising approaches and tools for mining architectural information. Specifically, the information presented in Table \ref{SuggestingApproachesToPractitioners} and Table \ref{SuggestingToolsToPractitioners} can guide practitioners know and adopt promising approaches and/or tools for mining certain types of architectural information from particular sources in order to support specific architecting activities. It is worth noting that we referred to the following factors (e.g., evidence levels, see Section \ref{ResultsOfRQ4}) to identify the most promising approaches and tools provided in Table \ref{SuggestingApproachesToPractitioners} and Table \ref{SuggestingToolsToPractitioners}. \textit{Adaptation in industrial practice (Level 6)}: Some tools are currently being utilized in industry settings, such as DV8. \textit{Evidence obtained from industrial studies (Level 5)}: An approach or tool has been evaluated using case studies in industry with practitioners. For example, Juergen et al. {[S53]} proposed an approach named Continuous Architectural Knowledge Integration (CAKI) that combines the continuous integration of internal and external architectural knowledge sources with enhanced semantic searching, reasoning, and personalization capabilities dedicated to large organizations. The approach was evaluated in an industrial setting with practitioners from Siemens. \textit{Extensive evaluation on non-trivial industrial software systems}: Some approaches and tools have been validated on a bunch of non-trivial industrial software projects. For example, in [S30],  Ghorbani et al. proposed an approach named DARCY for detecting architectural inconsistencies in Java software systems. The approach was evaluated on 38 open-source projects. 
  
\textbf{Suggestion 2}: \faHandORight \hspace{0.5mm} We suggest that practitioners could give high priorities to promising approaches and tools presented in Table \ref{SuggestingApproachesToPractitioners} and Table \ref{SuggestingToolsToPractitioners} when selecting approaches and tools for mining architectural information from software repositories.

\small
\begin{longtable}{p{5em}p{10.5em}p{7em}p{8em}p{2.9em}}
\caption{Promising approaches for mining architectural information} \label{SuggestingApproachesToPractitioners} \\\hline 
\textbf{Approach} & \textbf{Architectural information} & \textbf{Source and dataset} & \textbf{Supported architecting activities} & \textbf{Studies} \\\hline  
\endfirsthead
{{\bfseries }}\\
\hline
\endhead
\endfoot
\hline
\endlastfoot                               

{ARCADE}       
                                    & Architectural description (e.g., Architectural view)
                                    & VCS
                                    & Architectural recovery, Architectural maintenance and evolution
                                    & {[S46]}\\ \cline{1-5}  
{CAKI}       
                                    & Architectural solution (e.g., Architectural pattern, Architectural tactic), Architectural decision (e.g., Technology decision)
                                    & Mixed sources (VCS, Wiki, Q\&A sites, Online books, research articles, etc.)
                                    & Architecture Recovery
                                    & {[S53]}\\ \cline{1-5}

{Architecture Anti-pattern Detector}
                                    & Architectural technical debt (e.g., Architectural anti-pattern)
                                    & VCS, Issue tracking systems
                                    & Architecture maintenance and evolution
                                    & {[S64]}\\ \cline{1-5}                                     

{ArchDRH} 
                                    & Design relationship (e.g., Component-Component relationship)
                                    & VCS
                                    & Architecture recovery, Architecture maintenance and evolution
                                    & {[S101]}\\ \cline{1-5}
                                                                                                   
{Tactic Detector(TD)} 
                                    & Architectural solution (e.g., Architectural tactic)
                                    & VCS, Code search engines
                                    & Architecture reuse, Architecture implementation
                                    & {[S57]}\\ \cline{1-5}

{DARCY} 
                                    & Architectural technical debt (e.g., Architectural compliance issue), Architectural description (e.g., Architectural model)
                                    & VCS
                                    & Architecture conformance checking, Architecture understanding, Architecture maintenance
                                    & {[S30]}\\ \cline{1-5}                                                                      
\end{longtable}	
\begin{longtable}{p{3.5em}p{7em}p{5em}p{6em}p{6.9em}p{2.8em}}
\caption{Promising tools for mining architectural information} \label{SuggestingToolsToPractitioners} \\\hline 
\textbf{Tool} & \textbf{Architectural information} & \textbf{Source and dataset} & \textbf{Supported architecting activities} & \textbf{URL link} & \textbf{Studies} \\\hline  
\endfirsthead
{{\bfseries }}\\
\hline
\endhead
\endfoot
\hline
\endlastfoot

{Arcan} 
                                    & Architectural technical debt (e.g., Architectural smell)
                                    & Issue tracking systems (Code-based dataset)
                                    & Architectural maintenance and evolution
                                    & \url{https://essere.disco.unimib.it/wiki/arcan/} 
                                    & {[S88]}\\ \cline{1-6}

{DV8} 
                                    & Architectural technical debt (e.g., Architectural smell)
                                    & VCS (Code-based dataset)
                                    & Architectural maintenance and evolution
                                    & \url{https://archdia.com/}
                                    & [S94]\\ \cline{1-6}                                     


        
\end{longtable}	
\normalsize
\color{black}

\subsubsection{Developing tools to better support the approaches} 
From the analysis of the RQ4 results in Section \ref{Discussion_RQ4 and RQ2}, we can see that, for the proposed architectural information mining approaches (i.e., 95 approaches), still more than half of the approaches (i.e., 56 approaches) are not supported by tools. For instance, in {[S78]}, Kamal and Avgeriou proposed a manual approach for mining architectural pattern-architectural pattern relationships from books, research articles, and white papers, in order to assist architects in reusing architectural patterns during the architecting process. (Semi-)automating this approach with a dedicated tool can significantly improve the efficiency of architecting activities, such as architectural reuse (AReu), maintenance and evolution (AME).

\textbf{Suggestion 3}: \faHandORight \hspace{0.5mm} Future research may consider to examine the proposed mining approaches and develop tools to better support those approaches. For example, to enhance the applicability of the proposed approaches in practice, future work can implement those approaches as plugin tools that can be used within existing IDEs (e.g., Eclipse, PyCharm, Visual Studio) by adapting or following the techniques in prior studies, such as \cite{weinreich2012towards, harbo2024acsmt}. 

\subsubsection{Verifying the industrial needs for mining architectural information}
From the results of RQ4 (see Section \ref{ResultsOfRQ4}), we observed that most primary studies motivated their mining approaches to reduce the manual effort in mining architectural information. Such rationale can be reasonable in the context of large-scale collections of software artifacts. Furthermore, manually mining relevant architectural information from a large volume of software artifacts is not cost-effective. However, we have recorded no research studying what features the approaches or tools should possess nor what goals they should satisfy when mining architectural information from practitioners' perspectives. Thus, the current research mainly satisfies requirements coming from scholars rather than practitioners. This indicates that the current research is \textit{intuition-based} rather than \textit{practitioner-oriented}. These studies applied different types of approaches and tools to mine architectural information from software repositories without explicitly examining the practitioners’ needs, and it is not clear to what extent the approaches and tools satisfy the real practitioners’ needs. Although some existing studies provided evidence levels (e.g., evidence obtained from industrial studies [S53][S45][S55][S64]) of how practitioners found certain architectural information mining approaches and tools useful, more systematic research is necessary in such directions to understand practitioners’ needs. 

\textbf{Suggestion 4}: \faHandORight \hspace{0.5mm} Future research should plan to actively involve practitioners, for example, via interview sessions or analysis of their development practices, to understand what approaches and tools are needed in the industry. Such a study could address this problem from multiple perspectives, for example, what small, medium, and large software development organizations are interested in architectural information mining approaches and tools? Who in the organization would use the approaches or tools? Is mining architectural information manually ‘the real pain’ of practitioners? if so, how does ‘the pain’ manifest itself? Are any tasks obstructed? Is the problem generating additional costs? What architecting activities do the practitioners want to support with the approaches and tools? What features the approaches and tools should possess? Answering these questions could not only help to scope and justify future research directions on mining architectural information, but also provide insights into commercializing this research. Moreover, such knowledge will help to understand the actual use case scenarios of approaches and tools and to identify whether there is a misalignment between what state-of-the-art approaches and tools offer and what practitioners actually need.

\subsubsection{Open source benchmark architectural datasets for architecture research} 
The results of RQ5 report the challenges related to the quality and size of architectural datasets (e.g., lack of adequate training data {[S42]}). The quality of architecture-related datasets is important for architectural information mining approaches and tools. A significant portion of the research aimed to improve the quality of architectural datasets, including acquiring features, formulating data, cleaning data, and obtaining expert labeling of data. For instance, in the task of semi-automatic architecture information classification, the ground truth of the training data is crucial to the final results. The unstructured nature of the text and documents used makes the tasks of labeling data and selecting features a challenge. Regarding these problems, it is important to construct benchmark architectural datasets to facilitate research in the field of mining architectural information. A number of studies have worked on alleviating these problems by constructing large open-source architectural datasets from software repositories (e.g., \cite{zogaan2017automated, garcia2021constructing}) to support specific architectural research topics. However, sufficiently large-scale and quality architectural datasets for certain architecture research topics are not yet available. Moreover, the results of RQ5 report the challenges related to the limited applicability of the approaches to heterogeneous architectural datasets across software projects and artifact types. For example, an architectural information mining approach trained on a dataset from online Q\&A sites did not work well on an architectural dataset from GitHub \citep{mahadi2020cross}. As pointed out by \cite{canfora2015defect}, most software projects are heterogeneous and have various feature spaces and metric distributions. Certain context factors (e.g., size, domain, and programming language) may have a huge impact on data heterogeneity. 

\textbf{Suggestion 5}: \faHandORight \hspace{0.5mm} The software architecture community should work towards the construction of large-scale and quality benchmark architectural datasets (e.g., training datasets) for various architecture research topics. Recent studies experimented with active learning [S10] and semi-supervised leaning techniques \citep{naghdipour2023implications} to reduce the cost of annotating a large amount of data. Thus, these techniques and the like could be utilized in the construction of large-scale and quality benchmark architectural datasets. In addition, the software architecture community might address the limited applicability of architectural information mining approaches by accommodating the data heterogeneity issue in architectural datasets. Recently, transfer learning-based approaches have proven their power to deal with the data heterogeneity issue across software projects in several studies \citep{li2020automated}. Hence, future research could consider employing transfer learning-based approaches and resolve the challenges related to the limited applicability of architectural information mining approaches to heterogeneous architectural datasets. 

\subsubsection{Towards innovative approaches and tools to tackle the challenges in architectural information description}
The results of RQ5 also highlight the challenges related to architectural element descriptions in certain sources, including vagueness, incompleteness, and redundancy in architectural element descriptions, which may lead to, for example, poor performance of ML-based architectural information mining approaches. Such challenges are raised due to, for example, most software repositories containing unstructured data, such as unlabeled, vague, and noisy data \citep{hassan2008road}. For example, in {[S20]}, the authors clarified the vagueness in some sentences that describe software components in repositories, such as “\textit{That is a necessary component}”. This sentence is vague because it contains no explicit links to, for example, the source code entity (e.g., module, package) to which it refers. Another example is “\textit{Peer Authentication is supported on local sockets only}”. This sentence is vague and incomplete because additional context is needed since the terms “Peer Authentication” and “local sockets” are broad enough that they need association with a specific system or situation. Prior studies have proposed several approaches for dealing with vague and incomplete software development artifacts, such as extending incomplete API code snippets into compilable ones based on program analysis \citep{alon2020structural}, ML-based \citep{nguyen2019graph}, and DL-based \citep{ciniselli2021empirical} approaches. 

\textbf{Suggestion 6}: \faHandORight \hspace{0.5mm} The common way to deal with vague and incomplete software artifacts is to examine such artifacts’ features in terms of semantic analysis and understand the context those artifacts are used within the corpus or repository \citep{ciniselli2021empirical,liu2020multi}. Recently, ML and DL-based approaches have proven their power to capture semantic meaning from natural language in several studies \citep{ciniselli2021empirical, chen2020enhancing, wang2018deep}. Thus, in this regard, semantic meaning-based approaches [S65][S77][S82] using ML or DL techniques could be a starting point to be explored and deal with vagueness and incompleteness in architectural element descriptions when mining architectural information. For example, if the aim is to resolve vagueness and incompleteness in the descriptions of architectural patterns in certain corpus, such an approach could be developed to extract the context-sensitive features from this architectural pattern under investigation (e.g., the architectural components that constitute the architectural pattern under focus or investigation, as well as the relations between those architectural components and other architectural elements). These features can then be used to search and rank the patterns that are best matched with the current architectural pattern under investigation. When a pattern is chosen by a user (e.g., a developer), the approach would fill in the architectural elements based on that pattern with proper replacements of the corpus elements in the current context.
\section{Threats to Validity}\label{ThreatValidity}

In this section, we discuss the potential threats to the validity of our SMS, as well as the measures that we used to mitigate the threats according to the guidelines for identifying and mitigating threats to validity in software engineering secondary studies presented by \cite{ampatzoglou2019identifying}.

\subsection{Study Selection Validity}

Study selection validity is recognized as the major threat in SMSs during the early phases of the research. Therefore, in order to ensure that our search process has adequately identified all relevant studies, the primary studies that have been selected for inclusion were carefully checked following a well defined protocol based on strict guidelines \citep{petersen2015guidelines}. Specifically, to mitigate the risk of losing relevant studies, a search strategy was prepared and performed in our research. Search terms were constructed with the strings identified by checking the titles and keywords from relevant publications already known to the authors. However, we agreed that there may be relevant studies that were omitted, which affected the completeness of the retrieved results.
To further mitigate this threat, we searched in eight popular electronic databases (see Table \ref{ElectronicDatabases}) that publish software engineering research. Before the formal search, a pilot search was performed to enhance the suitability (e.g., the search terms) and quality of this SMS. Moreover, to ensure the completeness of the study search, we used the snowballing technique \citep{wohlin2014guidelines} to include any potentially relevant studies that were possibly missed out during the automatic search. After the set of primary studies were obtained, we proceeded to the study inclusion/exclusion phase, which is threatened by the possibility to exclude some relevant articles. To alleviate this threat, firstly, we defined a set of inclusion and exclusion criteria (see Table \ref{InclusionExculusionCriteria}) regarding the study selection. Also, the inclusion/exclusion criteria have been extensively discussed among the authors, so as to guarantee their clarity and prohibit misinterpretations. Secondly, before the formal study screening (manual inspection), to reach an agreement about the inclusion and exclusion criteria, a pilot study screening was performed whereby the first two authors randomly selected 100 studies (from the 19,588 retrieved studies) and checked them independently by following three rounds of study screening (see Section \ref{StudyScreening}). To measure the inter-rater agreement between the first two authors, we calculated the Cohen’s Kappa coefficient \citep{cohen1960coefficient} and got an agreement of 0.935. The results from the pilot study screening were checked and verified by the first two authors of this SMS, and they also discussed about the uncertain studies for reaching an agreement and reducing the risk that relevant studies were omitted.

\subsection{Data Validity}

The first potential threat to data validity is related to data extraction bias \citep{ampatzoglou2019identifying}. Several measures were taken to mitigate the bias of the researchers who conducted data extraction. First, to ensure the consistency of data extraction results, we discussed and formulated the data items by the first three authors of this study to reach a consensus on the content of the data to be extracted. In addition, before the formal data extraction, the first author conducted a pilot data extraction with 15 selected studies. The pilot data extraction results from the first author were checked by the second and third authors according to the description of each data item. Any disagreement was discussed and resolved together for reaching an agreement about the understanding of the data items (specified in Table \ref{DataExtraction}). The second potential threat to data validity is related to data synthesis \citep{ampatzoglou2019identifying}. The quality of data synthesis may affect the correctness of the answers to our five RQs (see Table \ref{ResearchQuestions}). Researchers may have their own understanding of data synthesis, for instance, the categorization of extracted data. To minimize personal bias, we performed a pilot data synthesis before the formal one. Specifically, the first two authors randomly chose 5 primary studies (from 104 selected studies). They independently read the full text of these 5 studies and encoded the extracted data items (see Table \ref{DataExtraction}) from those studies in order to answer the five RQs. To improve the reliability of the pilot data synthesis results, the first two authors held a meeting and followed a negotiated agreement approach~\citep{campbell2013coding} to compare the data encoding results, then discussed their disagreements, confusion, and uncertain judgments on the data encoding results in an effort to reconcile them and arrive at a final version of the pilot data synthesis results in which all the discrepancies have been resolved.

\subsection{Research Validity}

In this SMS, the penitential threats to research validity include being familiar with secondary studies, reliability and repeatability of the study, the selection of RQs, the research method used, and the generalizability of the study results \citep{ampatzoglou2019identifying}. However, we have excluded two possible threats to validity due to the nature of our study. First, the authors of this SMS are familiar with secondary studies, since they have been involved in a large number of secondary studies as authors and reviewers. Thus, no mitigation actions were necessary. Moreover, we believe that the followed review process ensures the reliability and safe replication of our study. First of all, we followed a rigorous process in this SMS  (e.g., pilot data extraction and synthesis) which can be easily reproduced by other researchers. Second, the fact that the data extraction was based on the opinion of three researchers can to some extent guarantee the elimination of bias, making the dataset reliable. 
However, some threats to research validity have been identified and mitigated. First, through discussion among the authors we set five RQs that can be holistically mapped to the goal of this SMS. This is clearly depicted by the mapping of each RQ to the research goal of this study (see Section \ref{GoalResearchQuestions}). 
Second, the selection of the research method (i.e., an SMS) is suitable for the goal of this study and no deviations from the guidelines (i.e., guidelines for conducting an SMS \citep{petersen2015guidelines}) have been performed. To ensure the generalizability of our results, we performed a broad search to collect and synthesize a wide range of studies from all sub-fields of software architecture, without any focus on specific architecting activities (e.g., architecture understanding and architecture synthesis). However, one possible threat to generalizability is related to the scope of literature search used in this study. Our search does not consider the grey literature, such as blogs (e.g., Martin Fowler's blog) that practitioners use to contribute to software architecture. To mitigate this threat, our research could be further enhanced by including the grey literature and more sources about architectural information that is being discussed by practitioners. 
\section{Related Work} \label{RelatedWork}
This work is an SMS in which we studied and summarized the primary studies reporting on mining architectural information from software repositories to support the development (i.e., architecting activities). We did not find any secondary study or survey that specifically discusses this topic. However, we checked a number of existing secondary studies and surveys because of the similarities between those works and ours in terms of the research setting. 

\cite{dkabrowski2022analysing} presented a comprehensive SLR within the field of app review analysis, covering 182 papers published between 2012 and 2020. This SLR classified app review analysis not only in terms of mined information and applied data mining techniques but also supported software engineering activities, such as requirements engineering, design, testing, and maintenance. The SLR also reports on the quality and results of empirical evaluation of existing techniques and identified important avenues for further research. 
\cite{tavakoli2018extracting} performed a literature survey on 34 studies published from 2011 to 2017. Specifically, they reviewed the application of ML and NLP techniques on extracting, mining, and classifying useful information from users’ feedback, to distinguish between requirements-relevant information and other types of users’ comments. The results of this survey can inform the development of more effective and intelligent app review mining techniques and tools. On the other hand, \cite{zhao2021natural} conducted an SMS that surveyed 404 primary studies relevant to Natural Language Processing for Requirements Engineering (NLP4RE) research. Their SMS not only provides a collection of the literature in NLP4RE but also establishes a structure to frame the existing literature through categorization, synthesis, and conceptualization of the main theoretical concepts and relationships that encompass both RE and NLP aspects. Moreover. Their work produces a conceptual framework of NLP4RE, which is used to identify research gaps and directions, highlight technology transfer needs, and encourage more synergies between the RE and NLP communities.
\cite{nazar2016summarizing} reviewed works (from January 2010 to April 2016) on summarization of various data sources, including bug reports, mailing lists, and developer forums, to extract requirements related information. In addition, they discussed the applications of summarization, i.e., what tasks at hand have been achieved through summarization. Next, their review presented tools that are generated for summarization tasks or employed during summarization tasks. Moreover, they presented different summarization evaluation methods employed in selected studies as well as other important factors that are used for the evaluation of generated summaries, such as adequacy and quality. \cite{casamayor2012mining} conducted a non-systematic survey on text mining and NLP in the field of model-driven design. Specifically, they investigated the application of text mining techniques and NLP on textual requirements to assist architecture design. Moreover, they analyzed and detailed intelligent text analysis techniques utilized in software engineering tasks across the software life-cycle in order to analyze the works that focused on automatically bridging the gap between requirements and architectures. 

The abovementioned studies are related to our work since they collected and synthesized primary studies on mining development information (e.g., system requirements) from software repositories to support the development, such as requirement engineering, testing, and maintenance. Nevertheless, our work is different from the aforementioned studies since our SMS does not focus on collecting and synthesizing the studies on mining one type of development information (e.g., system requirements). Moreover, none of the aforementioned studies has conducted a comprehensive SMS pertaining to studies on mining various types of architectural information from software repositories to support the development (i.e., architecting process) which is the focus of this study. Thus, we believe that our study complements the existing work on mining development information to support the development.
\section{Conclusions} \label{ConclusionFurtureWork}
In this SMS, we aimed to give a comprehensive overview of the current state of research on mining architectural information. Specifically, we investigated the mined architectural information, sources used, supported architecting activities, and approaches and tools employed, as well as the challenges faced when mining architectural information. To achieve this goal, we automatically searched the studies in seven electronic databases. In addition, we complemented the automatic search with the snowballing technique. We finally selected 104 primary studies published in 55 venues for further data extraction and synthesis to answer the five RQs. 

Our main results and findings show that various categories and subcategories of architectural information have been mined to support the development. The \textit{architectural description} is the most mined category of architectural information. Numerous sources have been used to mine architectural information, such as \textit{VCS} (e.g., GitHub), \textit{Online Q\&A sites} (e.g., Stack Overflow), \textit{Wiki}, and \textit{Issue tracking systems}. We identified eleven architecting activities that can be supported by the mined architectural information, where \textit{architecture understanding}, \textit{architecture maintenance and evolution}, and \textit{architecture recovery} are the top three most frequently supported architecting activities. Many approaches and tools have been proposed and employed to mine architectural information. Architectural information classification is the most frequently supported task by the approaches in mining architectural information. The challenges in mining architectural information are mainly derived from the descriptions of architectural elements, wherein vagueness or ambiguity in architectural element description is the prevalent challenge. Moreover, the reviewed studies report the challenges related to a limited number of quality datasets with an appropriate size available for architectural information mining approaches and tools. These challenges should receive more attention in future studies.  

Among the existing approaches and tools, we found that the approaches such as ARCADE proposed by Behnamghader et al. {[S46]}, CAKI proposed by Musil et al. {[S53]}, ArchDRH proposed by Cai et al. {[S101]}, Tactic Detector proposed by Mirakhorli et al. {[S57]} are promising architectural information mining approaches. On the other hand, tools including DV8 developed by Mo et al. {[S94]} and Arcan developed by Fontana et al. {[S99]} are promising architectural information mining tools. 

The results and findings of this SMS provide meaningful implications for both researchers and practitioners in the software architecture community. Strong industrial evidence is needed to assess the effectiveness of the proposed architectural information mining approaches and tools. In addition, dedicated approaches shall be proposed to tackle the reported challenges in mining architectural information, such as the vagueness or ambiguity in architectural element descriptions. Moreover, we encourage more research in mining architectural information to address the existing challenges, such as constructing a large-scale and quality architectural dataset for architectural information mining approaches and tools. Such a dataset could be used by other researchers during the evaluation of the proposed architectural information mining approaches and tools. Overall, our SMS will help practitioners understand and adopt the state-of-the-art research results, such as what approaches and tools should be adopted to mine what architectural information from what sources to support the development (i.e., architecting activities). It will also assist researchers to be aware of the potential challenges and remedies for the identified research gaps.

For the future research, we observed that some sources are less mined during the development and architects or developers are storing architectural information in a limited way or do not store it at all in certain repositories. Thus, we want to investigate the reasons behind these phenomena. Moreover, we plan to complement our SMS results by including the study search results from the grey literature, such as blogs, reports, and conduct a multi-vocal study in order to enrich the findings of this SMS.

\section*{Acknowledgements}
This work is partially sponsored by the National Natural Science Foundation of China (NSFC) under Grant No. 62172311 and 62176099, the Special Fund of Hubei Luojia Laboratory, the financial support from the China Scholarship Council, Shenzhen Polytechnic University with Grant No. 6022312043K, and State Key Laboratory for Novel Software Technology at Nanjing University with Grant No. KFKT2022B37.

\section*{Data Availability Statements}
The Supplementary Material of the current study is available in the Zenodo repository at \citep{dataset}.

\bibliographystyle{spbasic}
\bibliography{references}

\end{document}